\documentclass[11pt,a4paper,epsfig]{article}

\usepackage{amsfonts}
\usepackage{epsfig}

\title{\textbf{Changing shapes: adiabatic dynamics of composite solitary waves}}
\author{A. Alonso Izquierdo$^{(a)}$,
M.A. Gonzalez Leon$^{(a)}$ \\ M. de la Torre Mayado$^{(b)}$ and J.
Mateos Guilarte$^{(b)}$
\\ {\normalsize {\it $^{(a)}$ Departamento de Matematica
Aplicada}, {\it Universidad de Salamanca, SPAIN}}\\{\normalsize
{\it $^{(b)}$ Departamento de Fisica}, {\it Universidad de
Salamanca, SPAIN}}}
\date{}

\begin{document}

\maketitle

\begin{abstract}
We discuss the solitary wave solutions of a particular
two-component scalar field model in two-dimensional Minkowski
space. These solitary waves involve one, two or four lumps of
energy. The adiabatic motion of these composite non-linear
non-dispersive waves points to variations in shape.
\end{abstract}

\section{Introduction}

A solitary wave travels \lq\lq {\it without changing its shape,
size, or, speed}", \cite{Scott}. In this paper we shall deal with
solitary waves in relativistic scalar field theory in a
two-dimensional space-time. Besides the prototypes, the kink of
the non-linear Klein-Gordon equation and the sine-Gordon soliton
found in models where the scalar field has only one component,
solitary waves have also been discovered in systems with
two-component scalar fields, \cite{Raja1}-\cite{Rajaraman}.

In a deformed $O(2)$ Linear Sigma model, the celebrated
Montonen-Sarker-Trullinger-Bishop (MSTB) model, see \cite{Monto1},
\cite{Trulli1}, there exist solitary waves with two non-null field
components.  The model was first proposed by Montonen
\cite{Monto1}, and Sarker, Trullinger and Bishop \cite{Trulli1} in
the search for non-topological solitons. Nevertheless, in both
papers the existence of two kinds of solitary waves, respectively
with one and two non-null components, was noticed. Slightly later,
Rajaraman and Weinberg \cite{Raja1} proposed the trial-orbit
method to study the two-dimensional mechanical problem equivalent
to the search for static solutions. By these means, they found
that the TK1 -one-component topological- kink is given by a
straight line trajectory in field space, whereas the TK2
-two-component topological- kinks come from semi-elliptic
trajectories; they also enlarged the list by discovering NTK2
-two-component non-topological- kinks bound to elliptic orbits.
Through numerical analysis, \cite{Trulli2}, \cite{Trulli3},
Subbaswamy and Trullinger, looking for more exotic orbits,
observed the existence of a one-parametric family of NTK2,
including that previously discovered in \cite{Raja1}. Moreover,
they discovered an unexpected fact; the \lq\lq {\it kink energy
sum rule}": NTK2 energy is the same as the addition of TK1 and TK2
energies. Magyari and Thomas, \cite{Magy1}, realized that the
reason for the sum rule lies in the integrability of the
equivalent mechanical problem. Ito went further to show that the
mechanical problem giving the solitary waves as separatrix
trajectories in the MSTB model is not only integrable but also
Hamilton-Jacobi separable, \cite{Ito1}, and, moreover,
distinguished between stable and unstable solitary waves, see
\cite{Ito2}, by applying the Morse index theorem. The full Morse
Theory for the MSTB model was developed by one of us in \cite{J1}.

Only TK2 kinks are stable and genuine solitary waves in the sense
that they travel without distortion in shape, size, and, velocity.
In a model proposed by Bazeia-Nascimento-Ribeiro and Toledo
-henceforth the BNRT model, \cite{B3}, \cite{B5} - things are
different. In this case, both the TK1 and the TK2 kinks
-discovered in the papers quoted above- are stable and degenerate
in energy. In fact, they are distinguished members of a
one-parametric family of kinks found in \cite{Shifman}, all of
them degenerate in energy, but composite in a certain sense. Apart
from the center of the kink, there is a second integration
constant; for large values of the second integration constant $b$
solitary waves seem to be composed of two lumps of energy, whereas
for $b$ small they only show one lump. In References \cite{Mar1}
and \cite{Mar} we first unveiled the whole manifold of solitary
waves of this model and then showed that in this system solitary
waves can travel without dispersion, although changing the
relative position of the two lumps.

The distribution of the lumps is, however, completely symmetric
with respect to the center of mass. In this paper, we shall study
a similar system to the MSTB and BNRT models in this framework.
The model can be understood as the dimensional reduction of the
planar Chern-Simons-Higgs model, \cite{Wif}, to a line and zero
gauge field. The whole manifold of composite solitary waves was
described in Reference \cite{Aai2}. We shall again derive the
solitary wave solutions of the model called A in the Reference
cited, now using the Bogomol'ny method, see \cite{BPS}. This
strategy is better suited for describing the wave properties than
the solution of the Hamilton-Jacobi equation. Later, Manton's
adiabatic principle, designed to elucidate the slow-motion
dynamics of topological defects, \cite{Manton}, will be applied to
our solitary waves.

The organization of this paper is as follows: In Section \S .2 we
introduce the model and describe the spontaneous symmetry-breaking
scenario. We also identify two systems of first-order differential
equations that are satisfied by solitary waves using the
Bogomol'nyi procedure. In Section \S .3, the solitary wave
solutions of the first-order equations will be obtained. We shall
show that there exist two basic kink-shaped solitary waves and
that the rest of the kinks are configurations of two or four basic
kinks. The problem of the stability of these waves will be also
addressed in this Section; the double solitary waves are stable
whereas quadruple kinks are unstable. In Section \S .4, we apply
Manton's principle to study the adiabatic dynamics of solitary
waves as geodesic motion in their moduli space. There is no
problem with double solitary waves but we also analyze the motion
of quadruple lumps by pushing them in the direction of instability
in configuration space. Finally, we explore how the system react
to a small perturbation of the potential term, preserving the
basic kinks; the induced forces give rise to a remarkable bound
state of kinks, similar to the breather mode of sine-Gordon
theory.

\section{The model: quintic non-linear Klein-Gordon equation}

We shall focus on a (1+1)-dimensional relativistic scalar field
theory model, whose dynamics is governed by the action:
\begin{equation}
S=\int dy^2  \left\{ \frac{1}{2}  \partial_{\mu} \chi^{*}
\partial ^{\mu} \chi-U(\chi,\chi^{*})
\right\} \label{eq:act}
\end{equation}
and $\chi(y_\mu)=\chi_1(y_\mu)+i\chi_2(y_\mu): {\Bbb
R}^{1,1}\longrightarrow {\Bbb C}$ is a complex scalar field. We
choose $g_{00}=-g_{11}=1$ and $g_{12}=g_{21}=0$ as the metric
tensor components in Minkowski space ${\Bbb R}^{1,1}$, whereas
\[
U[\chi,\chi^{*}]=\textstyle\frac{\lambda^4}{4 m^2} \chi^{*} \chi
\left( \chi^{*} \chi -\frac{m^2}{\lambda^2}
\right)^2+\frac{\beta^2}{2} \chi_2^2 \left[ \chi^{*} \chi
-\frac{m^2}{\lambda^2} \left(1-\frac{\beta^2}{2 \lambda^2}
\right) \right]
\]
sets the non-linear interactions in (\ref{eq:act}). $\lambda$, $m$
and $\beta$ are coupling constants of inverse length dimension.
Introducing non-dimensional variables $\chi = \frac{m}{\lambda}
\phi$, $y_{\mu} = \frac{\sqrt{2}}{m} x_{\mu}$ and
$\frac{\beta^2}{\lambda^2} = \sigma^2$, the action functional
reads:
\[
S=\frac{m^2}{\lambda^2} \int d^2x \left[ \frac{1}{2}
\partial_\mu \phi^* \partial^\mu \phi -V(\phi,\phi^*)\right]
\]
\begin{equation}
V(\phi,\phi^{*})=\frac{1}{2} \, \phi^{*} \phi \, (\phi^{*}
\phi-1)^2 +\sigma^2 \phi_2^2 \left(\phi^{*} \phi-1
+\frac{\sigma^2}{2}\right) \qquad . \label{eq:a1}
\end{equation}
$- V(\phi,\phi^{*})$ -plotted in Figure 1a- is a semi-definite
negative polynomial expression of degree six which depends on the
unique classically relevant non-dimensional coupling constant
$\sigma$.

Although the potential that we have chosen seems to be very
awkward it is physically interesting for two reasons. First,
$V(\phi_1,\phi_2)$ is a deformation of $V_{{\rm
CSH}}(\phi)=\frac{1}{2} \, \phi^{*} \phi \,(\phi^{*} \phi-1)^2$
that appears in the celebrated Chern-Simons-Higgs model, see
\cite{Wif} and References quoted therein. The CSH model is a
(2+1)-dimensional gauge field theory for a complex scalar (Higgs)
field where the kinetic term for the gauge field in the Lagrangian
is the first Chern-Simons secondary class, whereas the Higgs
self-interaction is ruled by $V_{\rm CSH}$. A phenomenological
theory for the fractionary quantum Hall effect has been
established on the basis of a non-relativistic version of the CSH
model in \cite{Zh}. Dimensional reduction to (1+1)-dimensions at
zero gauge field leads to a class of systems including our model
because there are no Goldstone bosons on the line \cite{Co}; the
infrared asymptotic behavior of the quantum theory would require
the modification of $V_{CSH}$ in such a way that the zeroes become
a discrete set and massless particles are forbidden. The
deformation of our choice comply with this requirement and,
moreover, ensures integrability of the dynamical system to be
solved in the search for kinks. Second, we shall show in Section
\S3 that the Hamilton characteristic function, the \lq\lq
superpotential" $W^I(\phi)$ (\ref{eq:sup1}), is precisely the
potential of the MSTB model. This fact links both systems in a
hierarchal way.

The second term in (\ref{eq:a1}) breaks the $U(1)$ symmetry of
this model. The ${\Bbb G}={\Bbb Z}_2 \times {\Bbb Z}_2$ group,
generated by the reflections $\phi_1 \rightarrow -\phi_1$ and
$\phi_2 \rightarrow -\phi_2$ in the internal space ${\Bbb C}$,
leaves, however, our system invariant.

\begin{figure}[htbm]
\centerline{\epsfig{file=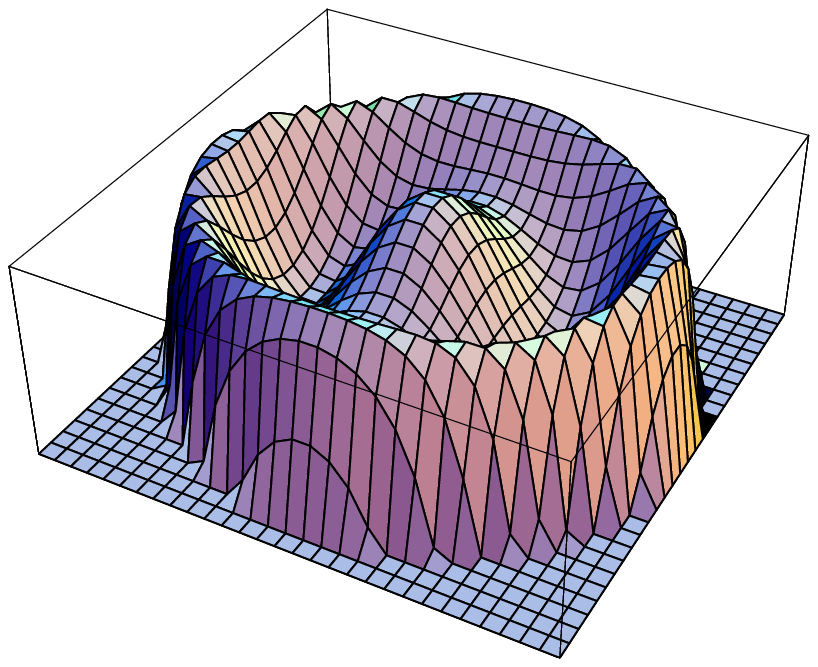,height=4cm}\hspace{1cm}\epsfig{file=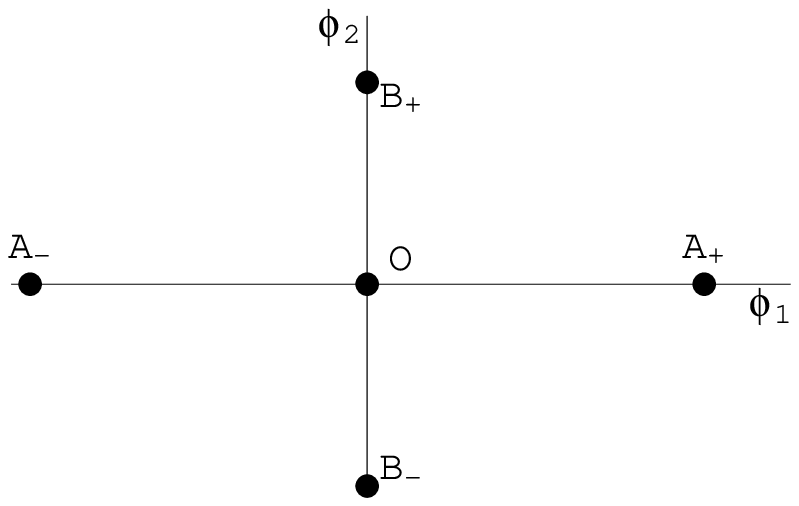,height=4cm}
} \caption{\it a) Potential term $-V(\phi,\phi^{*})$. b) Set of
zeroes ${\cal M}$ of $-V(\phi,\phi^{*})$.}
\end{figure}

The field equations form the following system of coupled
second-order PDE:
\begin{eqnarray}
\frac{\partial^2 \phi_1}{\partial t^2}-\frac{\partial^2
\phi_1}{\partial x^2}&=& -\phi_1 [1+3\phi_1^4+3\phi_2^4+
\phi_1^2(6\phi_2^2-4)+2(\sigma^2-2)\phi_2^2]  \label{eq:euler1}\\
\frac{\partial^2 \phi_2}{\partial t^2}-\frac{\partial^2
\phi_2}{\partial x^2}&=& -\phi_2
[3\phi_1^4+2\phi_1^2(3\phi_2^2+\sigma^2-2)+
(\phi_2^2+\sigma^2-1)(3\phi_2^2+\sigma^2-1)] \qquad .
\label{eq:euler2}
\end{eqnarray}
The PDE system (\ref{eq:euler1}), (\ref{eq:euler2}) is akin to the
non-linear Klein-Gordon equation with quintic -besides cubic-
non-linear terms. Initial, $\phi(t_0,x)$, $\frac{\partial
\phi}{\partial t}(t_0,x)$, and/or boundary conditions, $\phi(
t,\pm x_0)$, $\frac{\partial \phi}{\partial x}(t,\pm x_0)$, must
be chosen in order to search for physically relevant solutions. We
focus our attention on those solutions that can be interpreted as
lumps or extended particles: that is, solitary waves or kinks.
Recall the definition of a solitary wave, see e.g.
\cite{Rajaraman}: \lq\lq A solitary wave is a localized
non-singular solution of any non-linear field equation whose
energy density, as well as being localized, has space-time
dependence of the form: $\varepsilon(t,x)=\varepsilon(x-\mbox{v}
t)$, where v is some velocity vector".

\subsection{Structure of the configuration space}
The set of zeroes of $V(\phi_1,\phi_2)$ -maxima of
$-V(\phi_1,\phi_2)$- is:
\[
{\cal M}= \{\bar{\phi}^{A^\pm}=\pm 1 ; \, \bar{\phi}^{B^\pm}=\pm i
\bar{\sigma} ; \, \bar{\phi}^O =0 \} \qquad , \qquad
\bar{\sigma}=\sqrt{1-\sigma^2}
\]
see Figure 1b. These constant configurations are the static
homogeneous solutions of the PDE system (\ref{eq:euler1}),
(\ref{eq:euler2}) because they are critical points of
$V(\phi_1,\phi_2)$: $\frac{\partial
V}{\partial\phi_1}=0=\frac{\partial V}{\partial\phi_2}$. Note that
the reflection $\phi_1\rightarrow -\phi_1$ sends
$\bar{\phi}^{A^+}$ to $\bar{\phi}^{A^-}$ and vice-versa; they
belong to the same ${\Bbb G}$-equivalence class:
$\bar{\phi}^{A}=\{\bar{\phi}^{A^+},\bar{\phi}^{A^-}\}$. There are
another two ${\Bbb G}$-orbits between the constant solutions:
$\bar{\phi}^B=\{\bar{\phi}^{B^+},\bar{\phi}^{B^-}\}$ and
$\bar{\phi}^O$. The moduli space of homogeneous solutions- i.e.,
the set of zeroes of $V$ modulo the symmetry group- consists of
three points in ${\Bbb C}$: $\bar{\cal M}=\frac{{\cal M}}{{\Bbb
G}}= \{\bar{\phi}^A,\bar{\phi}^B,\bar{\phi}^O\}$. The degeneracy
of the homogeneous solutions -all of them have zero energy- causes
spontaneous breaking of the discrete symmetry. The constant
solution $\bar{\phi}^A$ breaks the symmetry under ${\Bbb G}={\Bbb
Z}_2\times {\Bbb Z}_2$ transformations of the action functional
(\ref{eq:act}) to the little group $H_1=\{e\} \times {\Bbb Z_2}$
generated by $\phi_2\rightarrow -\phi_2$. Simili modo, the
remaining symmetry group at $\bar{\phi}^B$ is the $H_2= {\Bbb
Z_2}\times \{e\}$ little group generated by $\phi_1\rightarrow
-\phi_1$, whereas the point $\bar{\phi}^O$ preserves the full
symmetry ${\Bbb G}$.

The configuration space of the system ${\cal C}=\{ \phi(t,x) \in
{\rm Maps}({\Bbb R}, {\Bbb C}) \hspace{0.3cm} / \hspace{0.3cm}
{\cal E}[\phi] < \infty \}$ is the set of maps from ${\Bbb R}$ to
${\Bbb C}$ for fixed time $t$ such that the energy functional
{\footnote {Strictly speaking the energy is
${m^3\over\lambda^2\sqrt{2}}{\cal E}$; ${\cal E}$, as given in
formula (\ref{eq:intro14}), is a non-dimensional quantity.}}
\begin{equation}
{\cal E}[\phi] =\int dx \left\{ \frac{1}{2}\, \left(\frac{\partial
\phi_1 }{\partial x}\right)^2 +\frac{1}{2}\, \left(\frac{\partial
\phi_2 }{\partial x}\right)^2+V(\phi_1,\phi_2) \right\}
\label{eq:intro14}
\end{equation}
is finite. Thus, every configuration in ${\cal C}$ must comply
with the asymptotic conditions:
\begin{equation}
\lim_{x \rightarrow \pm \infty} \phi(t,x) \in {\cal M}
\hspace{2cm} \lim_{x \rightarrow \pm\infty} \frac{\partial
\phi(t,x)}{\partial x} =0 \qquad . \label{eq:intro17}
\end{equation}
These asymptotic conditions play an important r$\hat{{\rm o}}$le
from a topological point of view. The configuration space is the
union of 25 topologically disconnected sectors:
\begin{eqnarray}
{\cal C}=&& \, \, \,\, {\cal C}^{AA}_{\pm\pm}\, \sqcup \,\, {\cal
C}^{BB}_{\pm\pm} \, \sqcup \,\, {\cal C}^{OO} \, \sqcup \,\, {\cal
C}^{AA}_{\pm\mp} \, \sqcup \,\, {\cal C}^{BB}_{\pm\mp}\nonumber
\\ && \, \sqcup \,\, {\cal
C}^{AB}_{\pm\pm} \, \sqcup \,\, {\cal C}^{BA}_{\pm\pm} \, \sqcup
\,\, {\cal C}^{AB}_{\pm\mp} \, \sqcup \,\, {\cal
C}^{BA}_{\pm\mp}\nonumber \\ && \, \sqcup \,\, {\cal
C}^{AO}_{\pm} \, \sqcup \,\, {\cal C}^{OA}_\pm \, \sqcup \,\,
{\cal C}^{BO}_{\pm} \, \sqcup \,\, {\cal
C}^{OB}_{\pm}\label{eq:conft}
\end{eqnarray}
The relation of this notation to the asymptotic conditions
(\ref{eq:intro17}) is self-explanatory. Three examples:
\begin{enumerate}
\item Sector ${\cal C}^{OO}$. Boundary conditions:
\[
\lim_{x\rightarrow -\infty} \phi(t,x)=\bar{\phi}^O \hspace{2cm} ,
\hspace{2cm} \lim_{x\rightarrow\infty} \, \phi (t,x)=\bar{\phi}^O
\]
\item Sector ${\cal C}^{AB}_{\pm\pm}$. Boundary conditions:
\[
\lim_{x\rightarrow -\infty} \phi(t,x)=\bar{\phi}^{A_\pm}
\hspace{2cm} , \hspace{2cm} \lim_{x\rightarrow\infty} \, \phi
(t,x)=\bar{\phi}^{B_\pm}
\]
\item Sector ${\cal C}^{BA}_{\pm\mp}$. Boundary conditions:
\[
\lim_{x\rightarrow -\infty} \phi(t,x)=\bar{\phi}^{B_\pm}
\hspace{2cm} , \hspace{2cm} \lim_{x\rightarrow\infty} \, \phi
(t,x)=\bar{\phi}^{A_\mp}
\]
\end{enumerate}
Because temporal evolution is continuous (a homotopy
transformation), the asymptotic conditions (\ref{eq:intro17}) do
not change with $t$ and the 25 sectors (\ref{eq:conft}) are
completely disconnected. Physically this means that a solution in
a sector cannot decay into solutions belonging to other different
sectors; it would cost infinite energy.

The five homogeneous solutions of
(\ref{eq:euler1})-(\ref{eq:euler2}) that are zeroes of $V$ belong
to the five sectors ${\cal C}^{AA}_{\pm\pm}$, ${\cal
C}^{BB}_{\pm\pm}$ and ${\cal C}^{OO}$ respectively. Small
fluctuations around one of them,
$\psi(t,x)=\bar{\phi}+\delta\psi(t,x)$, $\bar{\phi}\in{\cal M}$,
solve the linear PDE system:
\begin{equation}
\sum_{b=1}^2\left(\Box\delta_{ab}+M^2_{ab}(\bar{\phi})\right)\delta\psi_b(t,x)=0
\qquad , \qquad M^2_{ab}(\bar{\phi})=\frac{\partial^2
V}{\partial\phi_a\partial\phi_b}(\bar{\phi}) \label{eq:leuler}
\end{equation}
The solution of (\ref{eq:leuler}) via the separation of variables
-$\delta\psi_a(t,x)={\rm exp}[i\omega t]f_a^\omega(x)$- leads to
the spectral problem for the second order fluctuation -or Hessian-
operator:
\[
{\cal H}(\bar{\phi})\left(\begin{array}{c} f_1^\omega(x) \\
f_2^\omega(x)
\end{array}\right)=\left(\begin{array}{cc} -{d^2\over dx^2}+M^2_{11}(\bar{\phi}) & M^2_{12}(\bar{\phi}) \\
M^2_{21}(\bar{\phi}) & -{d^2\over dx^2}+M^2_{22}(\bar{\phi})
\end{array}\right)\left(\begin{array}{c} f_1^\omega(x) \\
f_2^\omega(x)
\end{array}\right)=\omega^2\left(\begin{array}{c} f_1^\omega(x) \\
f_2^\omega(x)
\end{array}\right)
\]
${\cal H}(\bar{\phi})$ is a diagonal matrix differential operator
for the three points of $\bar{\cal M}$ with a positive definite
spectrum; every constant solution belonging to ${\cal M}$ is
stable. Thus, there are three types of dispersive wave-packet
solutions, living respectively in one of these sectors. The
building blocks are plane waves with their dispersion laws
respectively determined by:
\[
\begin{array}{ccccc}
{M^2(\bar{\phi}^A)= \left(
\begin{array}{cc} 4 & 0 \\ 0 & \sigma^4 \end{array}
\right)} &  & {M^2(\bar{\phi}^B)= \left(
\begin{array}{cc} \sigma^4 & 0 \\ 0 & 4 \bar{\sigma}^4 \end{array}
\right)} &  & {M^2(\bar{\phi}^O)= \left(
\begin{array}{cc} 1 & 0 \\ 0 & \bar{\sigma}^4 \end{array}
\right)}
\end{array} \qquad .
\]
The two branches for each type are:

\[
\begin{array}{|cc|c|cc|} \hline &&&&\\[-0.3cm]  \bar{\phi}^A & & \bar{\phi}^B & &
\bar{\phi}^O \\ \hline &&&&\\[-0.3cm] \omega^2(k)=k^2+4 & &
\omega^2(k)=k^2+\sigma^4 & & \omega^2(k)=k^2+1 \\
\omega^2(q)=q^2+\sigma^4 & & \omega^2(q)=q^2+4\sigma^4 & &
\omega^2(q)=q^2+\bar{\sigma}^4 \\ \hline
\end{array}
\]

\subsection{Solitary waves from integrable dynamical systems}

Any other static solution of (\ref{eq:euler1})-(\ref{eq:euler2})
is a solitary wave that lives in one of the remaining twenty
topological sectors of the configuration space. The field profiles
can be either kink- or bell-shaped, and Lorentz invariance allows
the use of the center of mass system. The energy density and field
equations read:
\begin{equation}
\varepsilon(x)= {1\over
2}\left(\frac{d\phi_1}{dx}\right)^2+{1\over
2}\left(\frac{d\phi_2}{dx}\right)^2+V(\phi_1,\phi_2) \hspace{1cm}
; \hspace{1cm} {d^2\phi_a\over dx^2}={\partial
V\over\partial\phi_a} \quad , \quad a=1,2 \qquad .
\label{eq:seuler}
\end{equation}
The search for solitary waves is tantamount to the search for
finite energy static solutions $\phi(x)$ of the PDE system
(\ref{eq:euler1})-(\ref{eq:euler2}), which, in turn, reduces to
the ODE system (\ref{eq:seuler}). A Lorentz transformation sends
the static solution $\phi(x)$ to $\phi(t,x) =\phi \left(
\frac{x-{\rm v}t}{\sqrt{1-{\rm v}^2}} \right)$.

The ODE system (\ref{eq:seuler}) is nothing more than the
equations of motion for a two-dimensional mechanical system:
understand $(\phi_1,\phi_2)$ as the \lq\lq particle" coordinates;
$x$ as the \lq\lq particle" time, and $U=-V$ as the \lq\lq
particle" potential energy. We shall use a mixture of
Hamilton-Jacobi and Bogomol'nyi procedures to solve this system.
\begin{equation} V(\phi_1,\phi_2)=\frac{1}{2} \left(
\frac{\partial W}{\partial \phi^1} \right)^2+\frac{1}{2} \left(
\frac{\partial W}{\partial \phi^2} \right)^2 \label{eq:intro22}
\end{equation}
is the \lq\lq time"-independent Hamilton-Jacobi equation for the
mechanical system and zero \lq\lq particle" energy. Using the
Hamilton characteristic function $W$, the solution of
(\ref{eq:intro22}), we write the field theory potential energy for
static configurations $\grave{a}$ $la$ Bogomol'nyi:
\begin{eqnarray}
{\cal E}[\phi_1,\phi_2]&=&\int \! dx \, \frac{1}{2} \left[  \left(
\frac{d \phi^1}{dx}\right)^2+\left( \frac{d \phi^2}{dx}
\right)^2+ \left( \frac{\partial W}{\partial \phi^1}\right)^2
+\left( \frac{\partial W}{\partial \phi^2}\right)^2   \right] \nonumber \\
&=&\frac{1}{2} \int_{-\infty}^\infty dx \left[ \left( \frac{d
\phi^1}{dx}-\frac{\partial W}{\partial \phi^1} \right)^2 +\left(
\frac{d \phi^2}{dx}-\frac{\partial W}{\partial \phi^2} \right)^2
\right]+|\cal{T}| \qquad . \label{eq:bogo}
\end{eqnarray}
Here,
\begin{equation}
{\cal T} =\int_{-\infty}^\infty dx \left( \frac{d \phi^1}{dx}
\frac{\partial W}{\partial \phi^1}+\frac{d \phi^2}{dx}
\frac{\partial W}{\partial \phi^2}  \right)=\int \,\, dW
\label{eq:ecero}
\end{equation}
is a topological charge, depending only on the sector of the
configuration space if $W$ is well-behaved enough to apply Stokes
theorem:
\[
{\cal
T}=W(\phi_1(\infty),\phi_2(\infty))-W(\phi_1(-\infty),\phi_2(-\infty))\qquad
.
\]
In this case, an absolute minimum of the field theory energy
(\ref{eq:bogo}), in the corresponding topological sector, is
reached by the solutions of the following system of first-order
equations:
\begin{equation}
\frac{d \phi_1}{d x} =\frac{\partial W}{\partial \phi_1}
\hspace{1cm};\hspace{1cm} \frac{d \phi_2}{d x} =\frac{\partial
W}{\partial \phi_2} \qquad ; \label{eq:trabaz}
\end{equation}
(in (\ref{eq:bogo}) the squared terms are always positive and the
other is a topological constant). It is easy to check that
solutions of the first-order equations (\ref{eq:trabaz}) also
solve the second-order (Euler-Lagrange) equations
(\ref{eq:seuler}) and consequently (\ref{eq:euler1}) and
(\ref{eq:euler2}). The energy of these solutions depends only on
the topological sector of the solitary wave, and it saturates the
topological bound $\cal{E}[\phi_{\rm K}]=|{\cal T}|$. The quotient
of the first by the second first-order equations gives the \lq\lq
orbits":
\begin{equation}
\frac{\partial W}{\partial\phi_2}d\phi_1=\frac{\partial
W}{\partial\phi_1}d\phi_2 \qquad . \label{eq:orb}
\end{equation}
\begin{equation}
\left\{\frac{\partial
W}{\partial\phi_1}(\phi_1,\tilde{\phi_2}(\phi_1))\right\}^{-1}d\phi_1=dx
\hspace{0.5cm} , \hspace{0.5cm} \left\{\frac{\partial
W}{\partial\phi_2}(\tilde{\phi_1}(\phi_2),\phi_2)\right\}^{-1}d\phi_2=dx
\qquad , \label{eq:orbits}
\end{equation}
where $\tilde{\phi}_1(\phi_2)$ (respectively
$\tilde{\phi}_2(\phi_1)$) is a solution of (\ref{eq:orb}), provide
the \lq\lq time"-schedules of the \lq\lq particle". In fact, this
system of three equations would be also obtained in the
Hamilton-Jacobi framework setting all the separation constants
equal to zero.

\section{The variety of solitary waves}

Using elliptic coordinates,
\[
u(x)={1\over
2}\left(\sqrt{(\phi_1(x)+\sigma)^2+\phi_2^2(x)}+\sqrt{(\phi_1(x)-\sigma)^2+\phi_2^2(x)}\right)
\in (\sigma,+\infty)
\]
\[
v(x)={1\over
2}\left(\sqrt{(\phi_1(x)+\sigma)^2+\phi_2^2(x)}-\sqrt{(\phi_1(x)-\sigma)^2+\phi_2^2(x)}\right)
\in (-\sigma,\sigma) \qquad ,
\]
one can show that the equation (\ref{eq:intro22}) is separable for
our choice of $V(\phi_1,\phi_2)$. Thus, two independent solutions
for $W$ as a function of $u,v$ are found. Back in Cartesian
coordinates, these are:
\begin{equation}
W^I(\phi)=\frac{1}{4} \left[ (\phi_1^2+\phi_2^2-1)^2+2 \sigma^2
\phi_2^2 \right] \qquad , \label{eq:sup1}
\end{equation}
\begin{equation}
W^{II}(\phi)=\frac{1}{4} \, \sqrt{(\phi_1+\sigma)^2+\phi_2^2}  \,
\sqrt{(\phi_1-\sigma)^2+\phi_2^2} \,
(\phi_1^2+\phi_2^2+\sigma^2-2) \qquad . \label{eq:sup2}
\end{equation}
$W^I$ is a polynomial function of $\phi_1,\phi_2$ without
singularities and everything proceeds smoothly. The partial
derivatives of $W^{II}$, however, are undefined at the points
$(\phi_1=\pm\sigma ,\phi_2=0)$; it is uncertain how to interpret
this second solution of the Hamilton-Jacobi equation. Note also
that if a function $W(\phi)$ verifies formula (\ref{eq:intro22})
then $-W(\phi)$ also satisfies this condition. Replacing $W$ by
$-W$ merely flips the sign of the topological charge.

The system of first-order differential equations (\ref{eq:trabaz})
determined by the first solution $W^I$ is:
\begin{equation}
\left\{\begin{array}{l}
\displaystyle\frac{d \phi_1}{d x} =\phi_1 (\phi_1^2+\phi_2^2-1) \\[0.4cm] \displaystyle\frac{d
\phi_2}{d x} =\phi_2 (\phi_1^2+\phi_2^2-1+\sigma^2)
\end{array}\right. \qquad ,
\label{eq:edo1}
\end{equation}
whereas the flow of the gradient of the second solution $W^{II}$
is ruled by the first-order ODE system:
\begin{equation}\left\{
\begin{array}{l}
\displaystyle\frac{d \phi_1}{d x}
=\frac{\phi_1[\phi_1^4+\phi_2^4+\sigma^2+\phi_1^2(2\phi_2^2-1-\sigma^2)
+\phi_2^2(\sigma^2-1)]}{\sqrt{(\phi_1-\sigma)^2+\phi_2^2}\sqrt{(\phi_1+\sigma)^2
+\phi_2^2}}
\\[0.5cm] \displaystyle \frac{d \phi_2}{d x}
=\frac{\phi_2[\phi_1^4+\phi_2^4+\phi_1^2(2\phi_2^2-1) +\sigma^2
(\sigma^2-1)
+\phi_2^2(2\sigma^2-1)]}{\sqrt{(\phi_1-\sigma)^2+\phi_2^2}\sqrt{(\phi_1+\sigma)^2+
\phi_2^2}}
\end{array}\right. \qquad .
\label{eq:edo2}
\end{equation}

\subsection{Solitary waves: basic types}

In order to solve the ODE systems (\ref{eq:edo1}) and
(\ref{eq:edo2}) we start by looking for restrictions on $\phi$
that make both systems coincide. The idea is akin to the Rajaraman
trial-orbit method, see \cite{Rajaraman}: guess a trajectory that
joins two points in $\bar{\cal M}$ and minimizes ${\cal E}$. It
happens that there are two types of curves in ${\Bbb C}$ for which
(\ref{eq:edo1}) and (\ref{eq:edo2}) become identical; they provide
the basic solitary waves of the model.

\begin{enumerate}

\item \noindent {\sc K$_1^{BO}$}: We shall now try the condition
$\bar{\phi}_1=0$ in systems (\ref{eq:edo1}) and/or
(\ref{eq:edo2}). We find kink solutions joining the points $O$ and
$B$ in the moduli space:
\begin{equation}
\bar{\phi}(x;a)=(-1)^\alpha \,i \, \frac{\bar\sigma}{\sqrt{2}} \,
\sqrt{1\, +(-1)^\beta \, {\rm tanh} \, [\bar{\sigma}^2(x-a)] }
\hspace{1cm} \alpha,\beta=0,1 \label{eq:a5}
\end{equation}
Because the orbits stay on the straight line $\bar{v}=0$ in the
elliptic plane joining $B$ with $O$, we shall refer to these four
solutions {\footnote{The subscript refers to the number of lumps,
which is equal to the number of pieces of straight lines that form
the kink orbit in elliptic coordinates, see \cite{Aai2}.}}-here we
consider the parameter $a$ to be fixed- as K$_1^{BO}$. A solitary
wave connects the points $O$ and $B_+$ if $\alpha=\beta=0$ and the
points $O$ and $B_-$ if $\alpha=1,\beta=0$ -living respectively in
${\cal C}^{OB}_\pm$-. For the cases $\alpha=0,\beta=1$ and
$\alpha=\beta=1$, we find respectively the antikinks of the above
mentioned solutions. One of these solutions has been depicted in
Figure 2, in addition to its density energy and its orbit in the
internal complex plane. In Figure 2c, we see why these solitary
waves are interpreted as lumps of energy or extended particles.
The energy of these solutions is:
\[
{\cal E}[{\rm
K_1^{BO}}]=|T|=\left|W^I[\bar{\phi}^B]-W^I[\bar{\phi}^O]\right|=
\left|W^{II}[\bar{\phi}^B]-W^{II}[\bar{\phi}^O]\right|=
\left(\frac{1}{2}-\frac{\sigma^2}{2}\right)^2 \,\, .
\]

\begin{figure}[htb]
\centerline{\epsfig{file=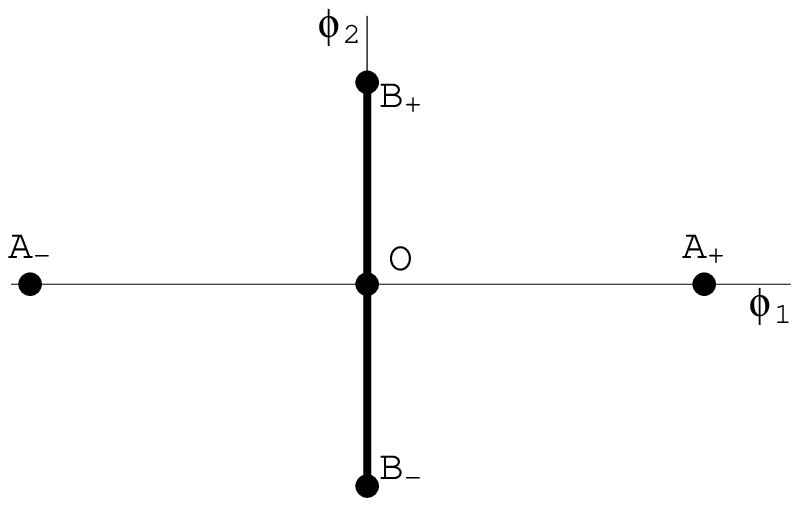,height=2.5cm}\hspace{0.5cm}
\epsfig{file=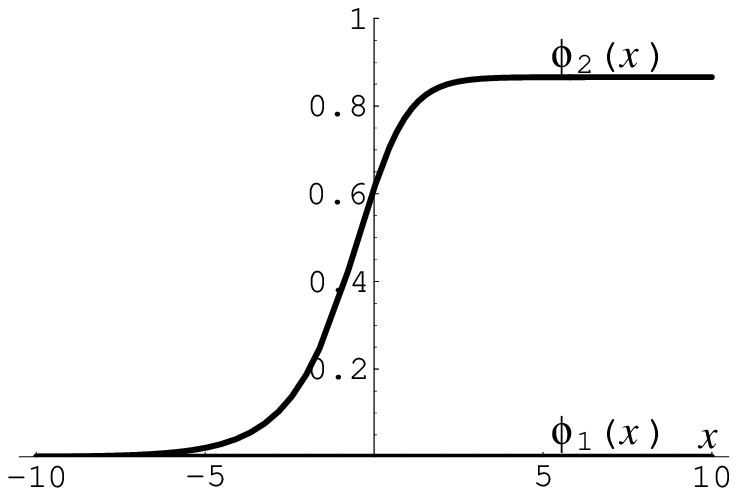,height=2.5cm} \hspace{0.5cm}
\epsfig{file=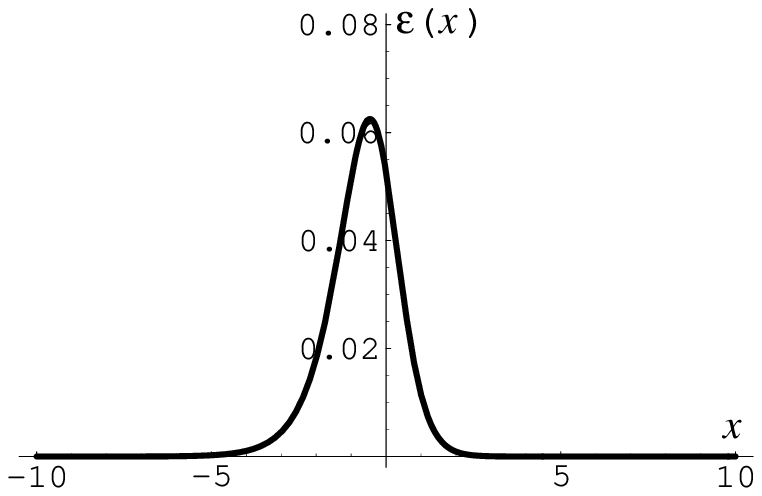,height=2.5cm}} \caption{\it {\rm
K}$_1^{BO}$ Kink: a) Orbit, b) Kink profiles and c) Energy
density.}
\end{figure}

\item \noindent {K$_1^{AB}$}: On the ellipse
$\bar{\phi}_1^2+\frac{\bar{\phi}_2^2}{1-\sigma^2}=1$ -$\bar{u}=1$
in the elliptic plane - (see Figure 3a), equations (\ref{eq:edo1})
and (\ref{eq:edo2}) also coincide. The following eight solitary
wave solutions are found:
\begin{equation}
\bar{\phi}(x;a)=\frac{(-1)^\alpha }{\sqrt{2}} \sqrt{ 1+{\rm tanh}
[ (-1)^\beta \; \sigma^2 \bar{x}] } \; + i \frac{(-1)^\gamma
\bar{\sigma}}{\sqrt{2}} \sqrt{1-{\rm tanh} [(-1)^\beta \;
\sigma^2\bar{x}] } \label{eq:a6}
\end{equation}
where $\alpha,\beta,\gamma=0,1\,,\, \bar{x}=x-a$ and $a$ is fixed.
All of them join the points $\bar{\phi}^A$ and $\bar{\phi}^B$ in
${\cal M}$ and thus belong to the ${\cal C}^{AB}$ sectors. We
shall denote these solitary waves depicted in Figure 3b as
K$_1^{AB}$ because their orbits lie on the straight line
$\bar{u}=1$ . If $\alpha=\gamma=0$, solution (\ref{eq:a6}) goes
from $B_+$ to $A_+$ when $\beta=0$ (see Figure 3b) and from $A_+$
to $B_+$ when $\beta=1$. The change in the value of $\alpha$ is
tantamount to a reflection $\phi_1\rightarrow -\phi_1$. Thus, for
$\alpha=1$, $\gamma=\beta=0$, (\ref{eq:a6}) is a solution running
from $B_+$ to $A_-$. Analogously, swapping the value of $\gamma$
in (\ref{eq:a6}) is equivalent to a reflection of the $\phi_2$
axis. For instance, (\ref{eq:a6}), with $\alpha=\beta=0$ and
$\gamma=1$, is the solution that starts from $B_+$ and ends in
$A_-$. Changing $\gamma$ kink and antikink are exchanged.

\begin{figure}[htb]
\centerline{\epsfig{file=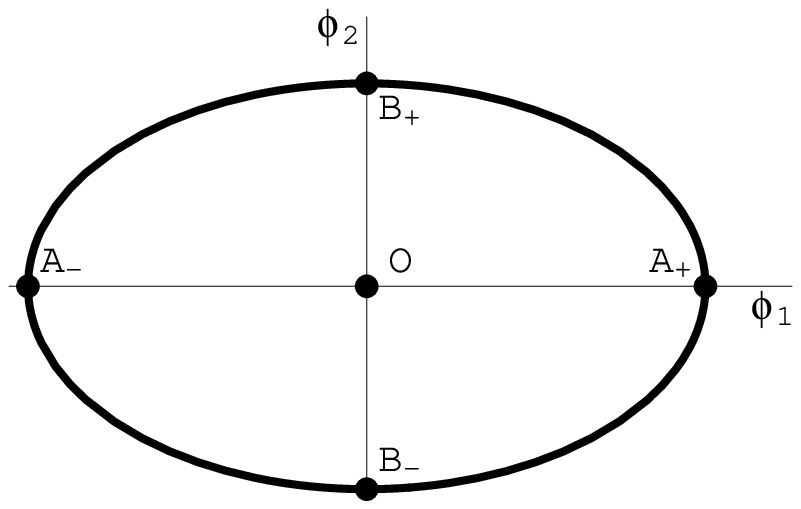,height=2.5cm}\hspace{0.5cm}
\epsfig{file=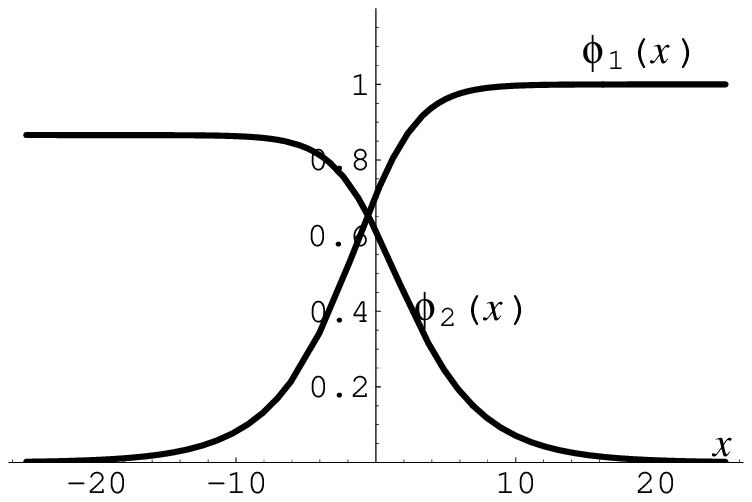,height=2.5cm} \hspace{0.5cm}
\epsfig{file=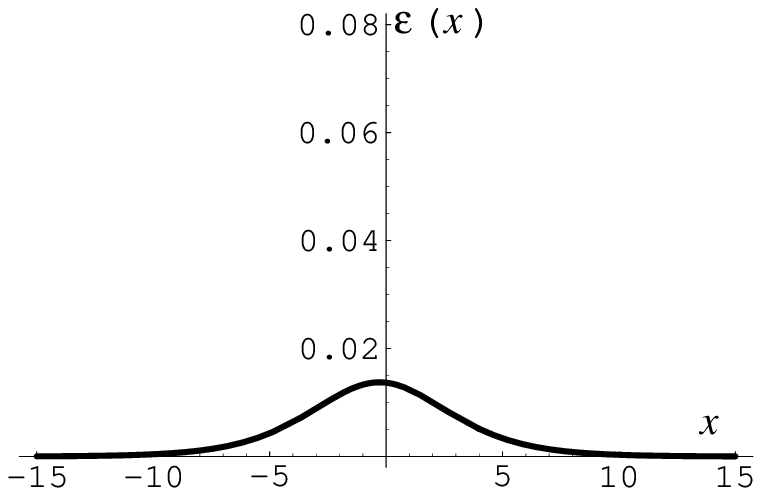,height=2.5cm}} \caption{\it {\rm
K$_1^{AB}$} Kink: a) Orbit, b) Kink profiles and c) Energy
density.}
\end{figure}

The energy density of these solitary waves is localized (see
Figure 3c) in a single lump (thus, K$_1^{AB}$). Their energy is:
\[
{\cal E}[{\rm K}_1^{AB}]=
\left|W^I[\bar{\phi}^B]-W^I[\bar{\phi}^A]\right|=
\left|W^{II}[\bar{\phi}^B]-W^{II}[\bar{\phi}^A]\right|=
\frac{\sigma^2}{2}-\frac{\sigma^4}{4} \qquad .
\]
\end{enumerate}

\subsection{Double solitary waves}

There exist many more solitary waves in this model than those
described in the previous sub-Section.

For example, we can try the straight line on the $\bar{\phi}_2=0$
axis joining the points $A$ and $O$ of the moduli space  ${\cal
M}$, see Figure 1b. Plugging this condition into both ODE systems,
(\ref{eq:edo1}) and (\ref{eq:edo2}), we find the common solution:
\begin{equation}
\bar{\phi}(x;a) = (-1)^\alpha \frac{1}{\sqrt{2}} \sqrt{1 \,
+(-1)^\beta \, {\rm tanh} (x-a)}  \hspace{1cm} \alpha,\beta=0,1
\qquad . \label{eq:a4}
\end{equation}
In fact, setting the solitary wave center at $a$, we have found
four solutions: (1) $\alpha=\beta=0$, the trajectory runs from the
point $O$ to $A_+$ as $x$ goes from $-\infty$ to $\infty$; it
belongs to ${\cal C}^{OA}_+$. (2) $\alpha=1,\beta=0$, the
trajectory starts from the point $O$ and arrives at $A_-$, living
in ${\cal C}^{OA}_-$. (3) $\alpha=0,\beta=1$ provides the solution
in ${\cal C}^{AO}_+$. (4) $\alpha=1,\beta=1$ leads to ${\cal
C}^{AO}_-$. The energy of these solutions is:
\[
{\cal E}[{\rm K_2^{AO}}]=|T|=|W^I[\bar{\phi}^A]-W^I[\bar{\phi}^O]|
=\frac{1}{4} \qquad .
\]
From the profile (\ref{eq:a4}) of the real component of the field
we see that these solitary waves are kink-shaped, interpolating
between the homogeneous solutions $\bar{\phi}^A$ and
$\bar{\phi}^O$. In Figures 4(center) and 5(center) we respectively
depict this solution and its energy density.

In the usual convention for kinks these solutions are of the type
TK1$^{AO}$, only one-Cartesian component being different from
zero. However, we denote these kinks as K$_2^{AO}$ because they
consist of two lumps, or, equivalently, their associated orbits in
the elliptic plane are formed by two straight lines: (a) from $A$
to $F$ along ${\bar v}=\pm\sigma$; (b) from $F$ to $O$ along
${\bar u}=\sigma$, where $F$ stands for the common foci of the
ellipses and hyperbolas defining the orthogonal system of
coordinates, see \cite{Aai2}.

There are many more solutions in the same topological sectors as
(\ref{eq:a4}). In \cite{Aai2} it is shown that system
(\ref{eq:edo1}) also separates into two independent first-order
ODE when written in elliptic coordinates. The general solution can
be found and translated back to Cartesian coordinates. The
solutions \lq \lq live" on the orbits:
\begin{equation}
\sigma^2 \, \bar{\phi}_2^2=b^2 (\sigma^2
\bar{\phi}_1^2)^{\bar\sigma^2} \left[ \bar\sigma^2
(1-\bar{\phi}_1^2)-\bar{\phi}_2^2 \right]^{\sigma^2} \qquad ,
\label{eq:0k2ao}
\end{equation}
which are solutions of equation (\ref{eq:orb}) for $W=W^I$, see
Figure 6(left). The explicit dependence on $x$ is given by solving
(\ref{eq:orbits}):
\begin{equation}
\bar{\phi}(x;a,b)=\frac{\bar{\sigma}}{\sqrt{\bar{\sigma}^2+
\sigma^2 e^{2 (x-a)}+b^2 e^{2 \sigma^2 (x-a)}}}+i \frac{ b
\bar{\sigma}}{\sqrt{b^2+\sigma^2 e^{2\bar\sigma^2
(x-a)}+\bar{\sigma}^2 e^{-2 \sigma^2 (x-a)}}} \qquad ,
\label{eq:tk2ao}
\end{equation}
where $a,b\in{\Bbb R}$ are two real integration constants.
Constant $a$ sets the center of the solitary wave. The meaning of
$b$ is two-fold. On one hand, $b$ parametrizes the family of
orbits (\ref{eq:0k2ao}) linking $\bar{\phi}^A$ and $\bar{\phi}^O$,
see Figure 6(left). On the other hand, $b$ enters the solitary
wave profile (\ref{eq:tk2ao}), shown in Figure 4 for $a=0$ and
several values of $b$. For high values of $|b|$, the real
component of the field has the shape of a kink but the imaginary
component is bell-shaped. When $|b|$ diminishes, the imaginary
component diminishes until it reaches zero for $b=0$.
\begin{figure}[htb]
\centerline{\epsfig{file=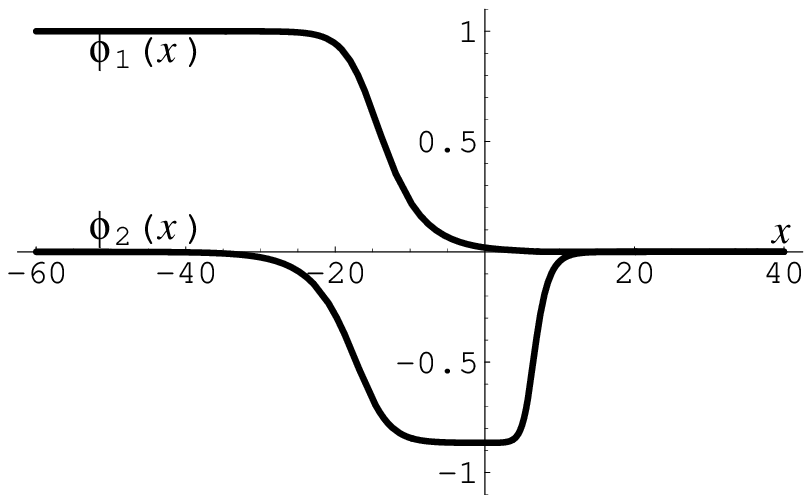,height=1.8cm}
\epsfig{file=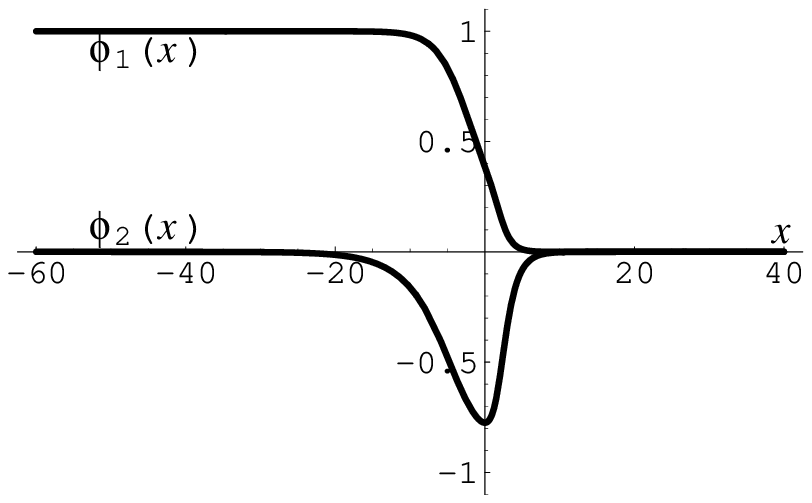,height=1.8cm}
\epsfig{file=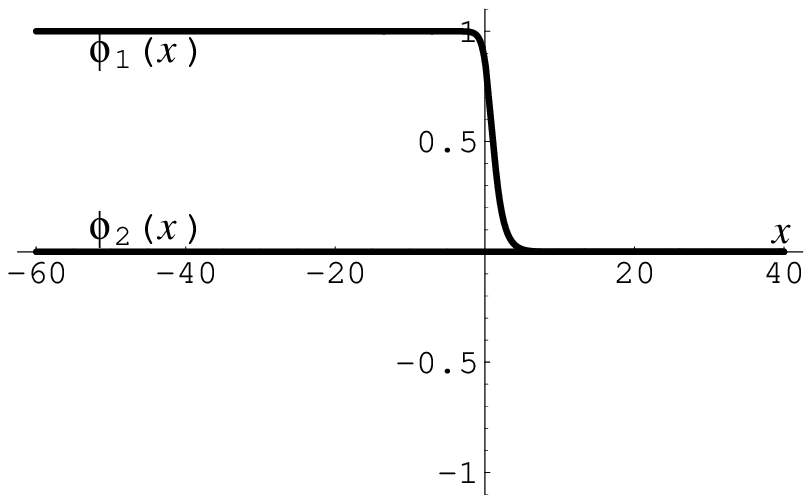,height=1.8cm}
\epsfig{file=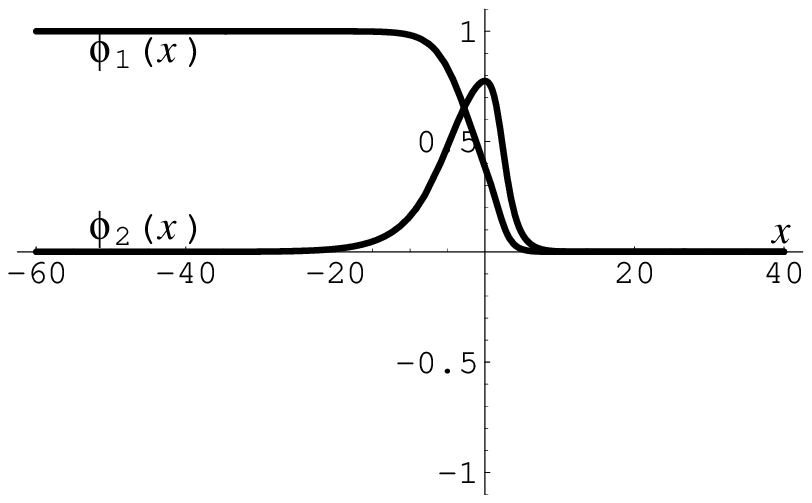,height=1.8cm}
\epsfig{file=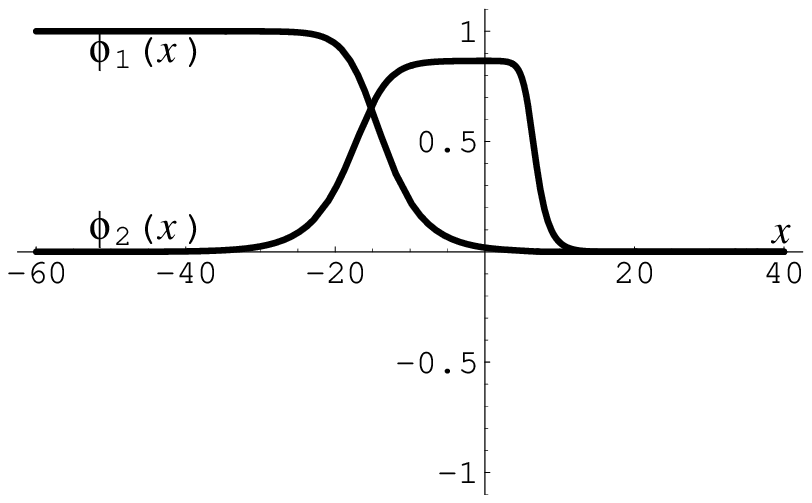,height=1.8cm}} \caption{\it Members
of the family {\rm K$_2^{AO}(b)$}: from left to right $b=-45, -2,
0, 2, 45$.}
\end{figure}
A parallel plot of the energy density for the same values of $b$
shows two energy lumps for high values of $|b|$ and only one lump
for $|b|$ close to zero, see Figure 5.
\begin{figure}[htb]
\centerline{\epsfig{file=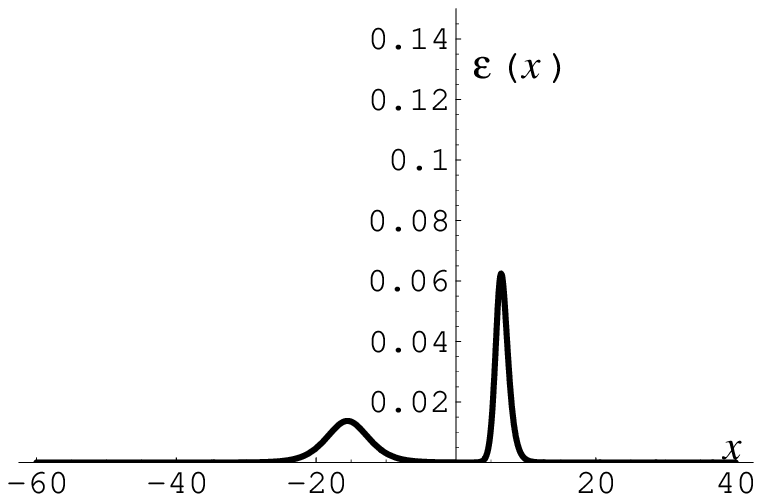,height=1.8cm}
\epsfig{file=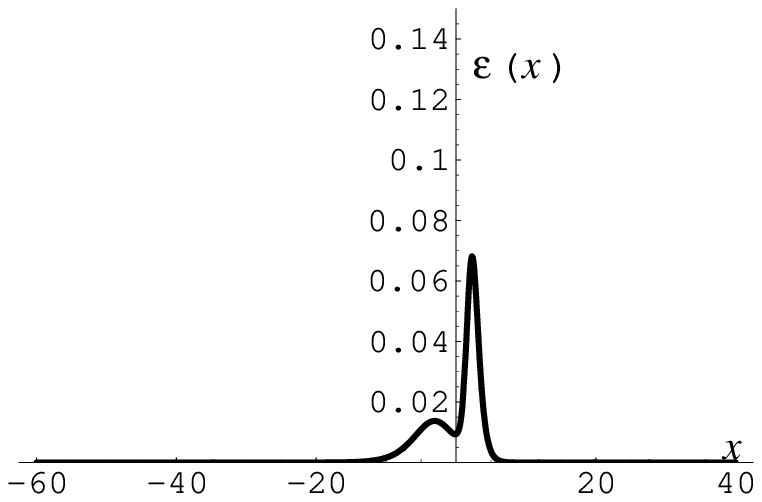,height=1.8cm}
\epsfig{file=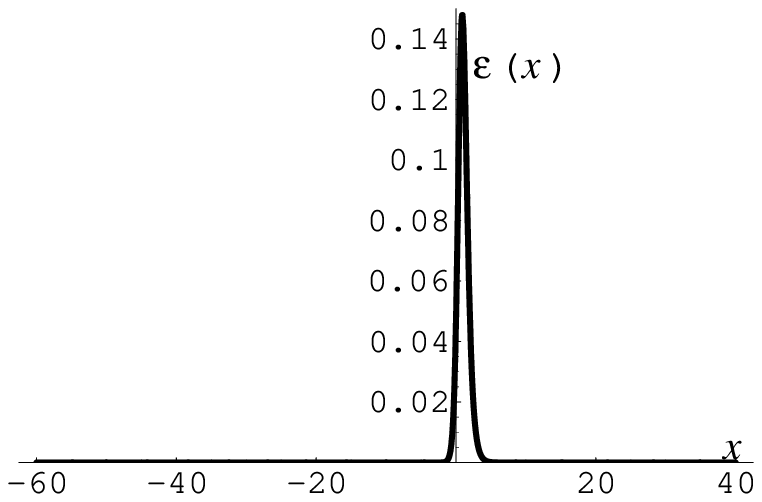,height=1.8cm}
\epsfig{file=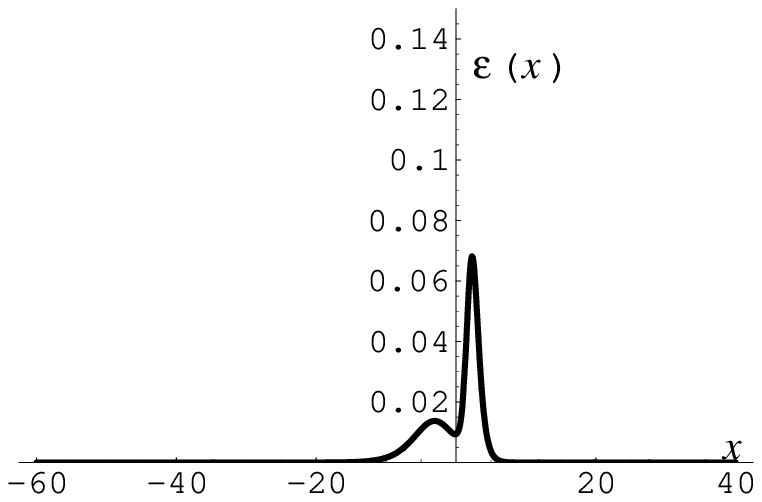,height=1.8cm}
\epsfig{file=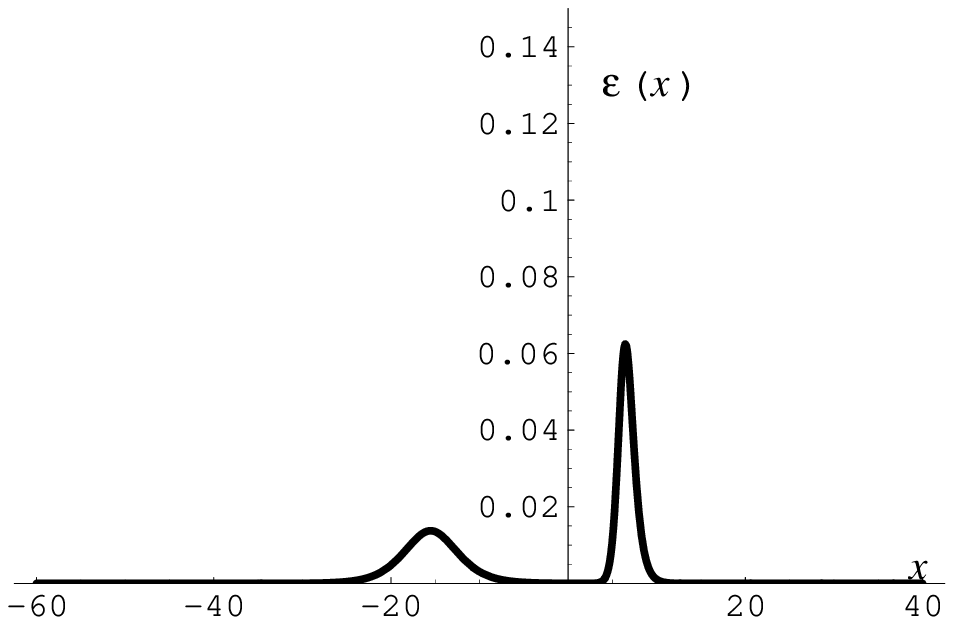,height=1.8cm}} \caption{\it Energy
density of the previous kinks: from left to right $b=-45, -2, 0,
2, 45$.}
\end{figure}
We denote these solitary waves as K$_2^{AO}(b)$ and realize that
the K$_2^{AO}(b)$ kinks seem to be made from the two basic kinks
K$_1^{BO}$ and K$_1^{AB}$ whereas K$_2^{AO}$=K$_2^{AO}(0)$. Both
facts are confirmed by the following tests: (1) the $|b|=\infty$
limit in (\ref{eq:0k2ao}) is only compatible with
$\bar{\phi}_1^2+\frac{\bar{\phi}_2^2}{\bar{\sigma}^2}=1$ and
$\bar{\phi}_1=0$ whereas the $b=0$ limit provides us with the
condition $\bar{\phi}_2=0$. (2) There is a new kink energy sum
rule for all $b$:
\begin{equation}
{\cal E}[{\rm
K_2}^{AO}(b)]=|W^I(\bar{\phi}^A)-W^I(\bar{\phi}^O)|={\cal E}[{\rm
K_1}^{BO}]+{\cal E}[{\rm K_1}^{AB}]=\frac{1}{4} \label{eq:sumrk2}
\end{equation}
As a curiosity, for the values of the coupling constant
$\sigma^2=\frac{1}{2}$ and $b$-parameter $b=\pm 1$ the
K$_2^{AO}(\pm 1)$ kinks  live on semi-circles of radius
$\frac{1}{2}$ with their centers on the points $(\pm
\frac{1}{2},0)$, i.e., $\left( \bar{\phi}_1\pm
\textstyle\frac{1}{2} \right)^2+\bar{\phi}_2^2 =\left(
\textstyle\frac{1}{2} \right)^2$. The profile is:
\[
\bar{\phi}(x;a)=\pm \textstyle\frac{1}{2} \left(1+\tanh
\textstyle\frac{(x-a)}{2} \right) \pm  2 \, i\, {\rm sech}\,
\textstyle\frac{(x-a)}{2}
\]

\begin{figure}[htbp]
\centerline{\epsfig{file=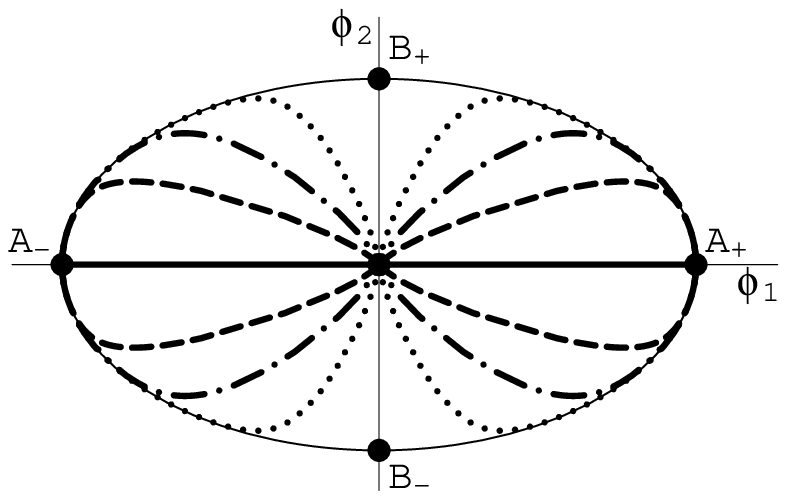,height=3.5cm}\hspace{1cm}
\epsfig{file=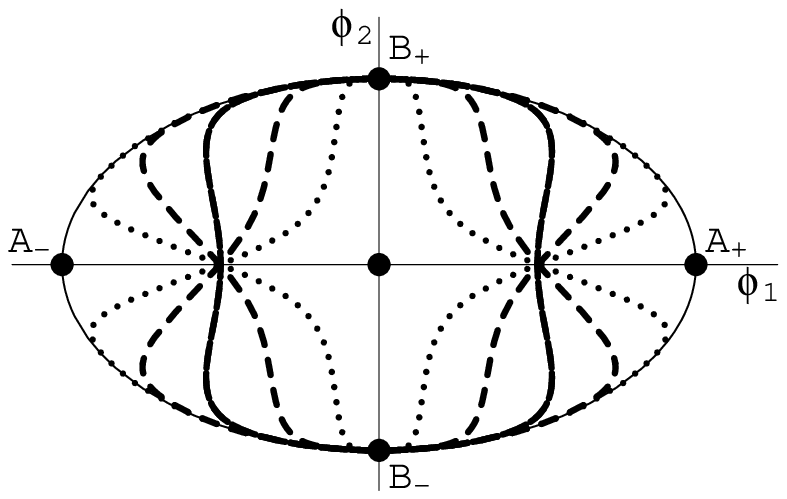,height=3.5cm}} \caption{\small \it
(left) Orbits for {\rm K}$_2^{AO}$ with $b=2$ (dotted line), $b=1$
(dash-dotted line), $b=0.5$ (broken line) and $b=0$ (solid line).
(right) {\rm K}$_4^{BB}$ kinks with $b=7$ (dotted line), $b=3$
(broken line) and $b=0$ (solid line).}
\end{figure}

\subsection{Quadruple solitary waves}
In Reference \cite{Aai2} we solved (\ref{eq:edo2}) again profiting
from the fact that this coupled ODE system becomes two ordinary
uncoupled differential equations in elliptic variables. Thus, it
is possible to obtain the equations of the orbits and
time-schedules explicitly. Translating the results back to
Cartesian coordinates, we found the complicated expressions:
\begin{eqnarray}
\left[C(\bar{\phi_1},\bar{\phi}_2)+D(\bar{\phi}_1,\bar{\phi}_2)-\sigma^2\right]
\left[(C(\bar{\phi}_1,\bar{\phi}_2)-D(\bar{\phi}_1,\bar{\phi}_2))^{\bar{\sigma}^2}
-C(\bar{\phi}_1,\bar{\phi}_2)+D(\bar{\phi}_1,\bar{\phi}_2)\right]&&\nonumber\\
=b^2
\left[\sigma^2-C(\bar{\phi_1},\bar{\phi}_2)+D(\bar{\phi}_1,\bar{\phi}_2)\right]
\left[(C(\bar{\phi}_1,\bar{\phi}_2)+D(\bar{\phi}_1,\bar{\phi}_2))^{\bar{\sigma}^2}
-C(\bar{\phi}_1,\bar{\phi}_2)-D(\bar{\phi}_1,\bar{\phi}_2)\right]&&
\label{eq:horb1}
\\ \left[\sigma^2-C(\bar{\phi_1},\bar{\phi}_2)+D(\bar{\phi}_1,\bar{\phi}_2)\right]
\left[(C(\bar{\phi}_1,\bar{\phi}_2)+D(\bar{\phi}_1,\bar{\phi}_2))^{\bar{\sigma}^2}
-C(\bar{\phi}_1,\bar{\phi}_2)-D(\bar{\phi}_1,\bar{\phi}_2)\right]&&
\nonumber \\
=b^2
\left[C(\bar{\phi_1},\bar{\phi}_2)+D(\bar{\phi}_1,\bar{\phi}_2)-\sigma^2\right]
\left[(C(\bar{\phi}_1,\bar{\phi}_2)-D(\bar{\phi}_1,\bar{\phi}_2))^{\bar{\sigma}^2}
-C(\bar{\phi}_1,\bar{\phi}_2)+D(\bar{\phi}_1,\bar{\phi}_2)\right]\,\,&&
\qquad , \label{eq:horb2}
\end{eqnarray}
where
\[
C(\bar{\phi}_1,\bar{\phi}_2)=\frac{1}{2}(\bar{\phi}_1^2+\bar{\phi}_2^2+\sigma^2)
\quad , \quad
D(\bar{\phi}_1,\bar{\phi}_2)=\frac{1}{2}\sqrt{(\bar{\phi}_1^2+\bar{\phi}_2^2)^2+2\sigma^2(\bar{\phi}_2^2
-\bar{\phi}_1^2)^2+\sigma^4} \quad ,
\]
as solutions of equation (\ref{eq:orb}) for $W=W^{II}$, see Figure
6 (right).

Also, we showed that
\begin{eqnarray}
\bar{\phi}_1(x;a,b)&=&\frac{(-1)^\alpha\sigma
(1+e^{2\bar{\sigma}^2
((x-a)+b\sigma^2)})}{\sqrt{\sigma^2+\bar{\sigma}^2
e^{2(x-a)}+e^{2\bar{\sigma}^2 ((x-a)+b\sigma^2)}}\sqrt{1+\sigma^2
e^{2\bar{\sigma}^2 ((x-a)+b\sigma^2)}+\bar{\sigma}^2
e^{-2\sigma^2 ((x-a)-b\bar\sigma^2)}}} \,\,\,\nonumber \\
\bar{\phi}_2(x;a,b)&=&\frac{\sigma \bar{\sigma}^2
e^{b\sigma^2}(e^{2(x-a)}-1)}{\sqrt{\sigma^2+\bar{\sigma}^2
e^{2(x-a)}+e^{2\bar{\sigma}^2
((x-a)+b\sigma^2)}}\sqrt{e^{2\sigma^2 ((x-a)+b\sigma^2)}+\sigma^2
e^{2((x-a)+b\sigma^2)}+\bar{\sigma}^2 e^{2b \sigma^2}}}\,\,\,\,
\label{eq:tk4bb}
\end{eqnarray}
solve the second-order equations of motion.

Besides the continuous parameters $a,b\in{\Bbb R}$, the family of
solitary waves (\ref{eq:tk4bb}) depends on the discrete value of
$\alpha=0,1$. All the solitary waves in the family
(\ref{eq:tk4bb}) belong to the sector ${\cal C}_{-+}^{BB}$; their
associated orbits go from $\bar{\phi}^{B_-}$ to $\bar{\phi}^{B_+}$
when $x-a$ goes from $-\infty$ to $+\infty$. When $x=a$ the foci
$\phi^{F_\pm}=\pm\sigma$ of the ellipse are respectively reached
for $\alpha=0$ and $\alpha=1$ {\footnote{Solitary waves of the
family (\ref{eq:tk4bb}) with $\alpha=0$ always remain in the
half-plane $\phi_1>0$. If $\alpha=1$, the solitary waves live in
the half-plane $\phi_1<0$.}}. The foci are the zeroes of
$D(\bar{\phi}_1,\bar{\phi}_2)$; moreover, the half-orbit equations
(\ref{eq:horb1}) and (\ref{eq:horb2}) are satisfied at these
points by all values of $b$, $0=b^2\cdot 0$. Therefore, an
infinite number of these trajectories arrives in and leaves from
the foci of the ellipse at the \lq\lq instant" $x=a$, see Figure 6
(right) and note that
\[
\frac{d\bar{\phi}_2}{d\bar{\phi}_1}=\frac{\bar{\phi}_2(\sigma^4+(\bar{\phi}_1^2
+\bar{\phi}_2^2-1)(\bar{\phi}_1^2
+\bar{\phi}_2^2)+\sigma^2(2\bar{\phi}_2^2-1))}{\bar{\phi}_1(\sigma^2(-\bar{\phi}_1^2
+\bar{\phi}_2^2+1)+(\bar{\phi}_1^2
+\bar{\phi}_2^2-1)(\bar{\phi}_1^2 +\bar{\phi}_2^2))}
\]
is indeterminate at $\phi^{F_\pm}=\pm\sigma$, although taking the
limit we obtain $\frac{d\bar{\phi}_2}{d\bar{\phi}_1}=\pm
\frac{2e^{\mp 2\sigma^2 \bar\sigma^2 b}}{1-e^{\mp 2\sigma^2
\bar\sigma^2 b}}$ at the foci.

Related to this fact, we stress the following subtle point: the
solitary wave family (\ref{eq:tk4bb}) does not strictly solve the
first-order ODE system (\ref{eq:edo2}) because the associated
orbits satisfy (\ref{eq:horb1}) if $x\in (-\infty,a)$ and
(\ref{eq:horb2}) if $x\in (a,+\infty)$. Taking the
$b^2\rightarrow\infty$ limit in (\ref{eq:horb1}), we find the four
K$_1^{AB}$ kinks and two times the $AF$ pieces of the K$_2^{AO}$
orbits; all the orbits in (\ref{eq:horb1}) have cuspidal points at
the foci. In the $b^2\rightarrow\infty$ limit of (\ref{eq:horb2}),
the two K$_1^{BO}$ kinks are obtained twice, together with the
$FO$ pieces of the K$_2^{AO}$ orbits; curves in this family also
have cuspidal points at the foci. Smooth trajectories require the
glue of half the orbits in (\ref{eq:horb1}) to their complementary
halves in (\ref{eq:horb2}). This means that these solitary waves
are solutions of (\ref{eq:edo2}) for $-W^{II}$ if $x\in
(-\infty,a)$ and properly for $W^{II}$ if $x\in (a,+\infty )$.

Nevertheless, they are bona fide solitary wave solutions of the
second-order field equations although the Stokes theorem apply
only piece-wise to the integration of $dW^{II}$ along these
orbits. The energy of solitary waves of this type is not a
topological quantity:
\begin{equation}
{\cal
E}(K_4^{BB}(b))=|W^{II}(\bar{\phi}^{B_-})-W^{II}(\phi^{F_\pm})|
+|W^{II}(\phi^{F_\pm})-W^{II}(\bar{\phi}^{B_+})|={1\over 2} \qquad
. \label{eq:qen}
\end{equation}
Because of the energy sum rule,
\begin{equation}
{\cal E}(K_4^{BB}(b))={\cal E}(K_1^{BO})+{\cal E}(K_2^{OA})+{\cal
E}(K_1^{AB})=2{\cal E}(K_1^{AB}(b))+2{\cal E}(K_1^{BO})\qquad ,
\label{eq:srdqen}
\end{equation}
we call these solitary waves, composed of four basic kinks,
K$_4^{BB}$. Replacing $x$ by $-x$ in (\ref{eq:tk4bb}), similar
solitary waves are found -K$_4^{BB}$ antikinks- living in ${\cal
C}^{BB}_{+-}$.

\begin{figure}[htb]
\centerline{\epsfig{file=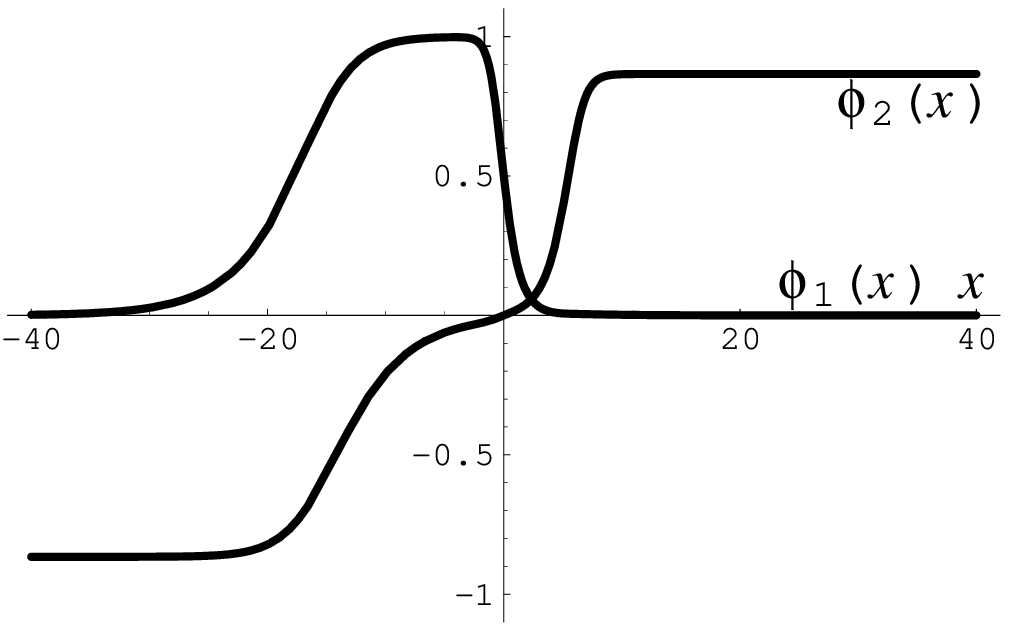,height=1.8cm}
\epsfig{file=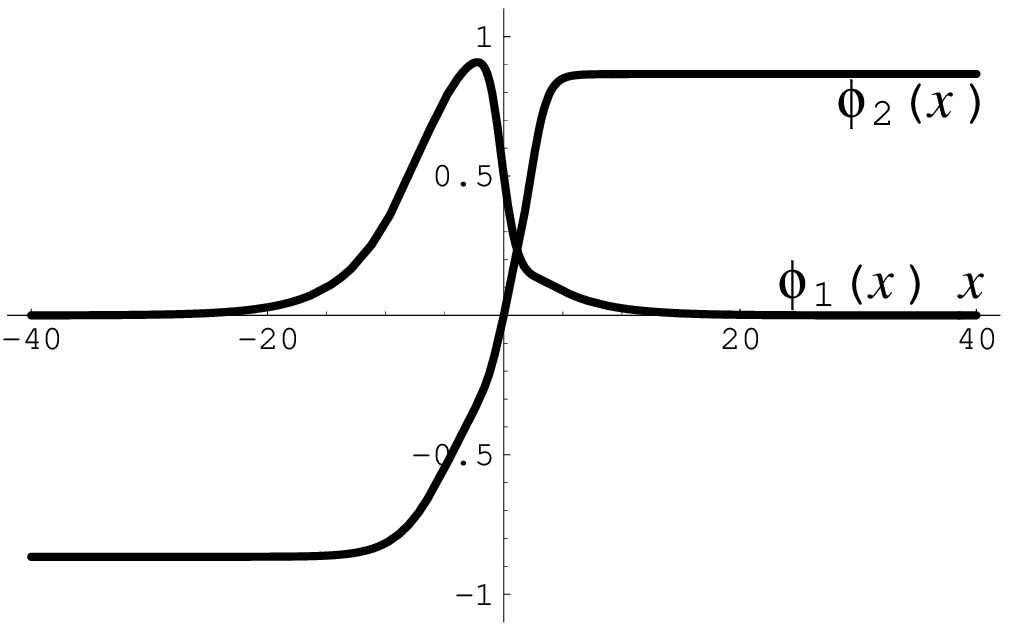,height=1.8cm}
\epsfig{file=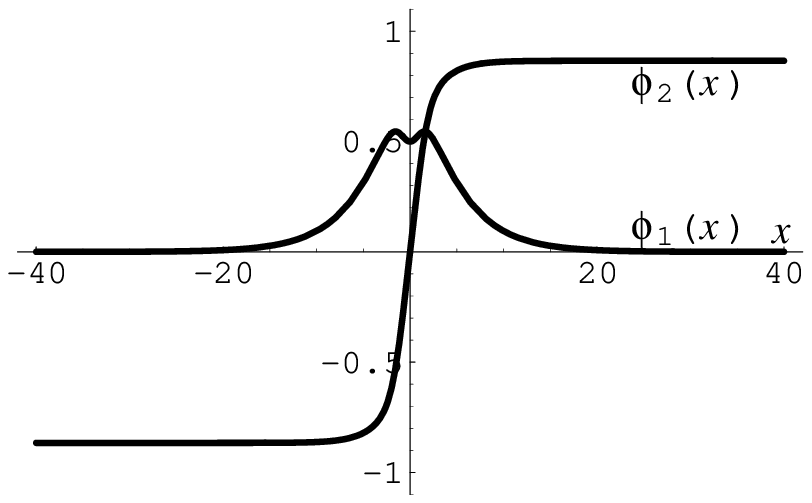,height=1.8cm}
\epsfig{file=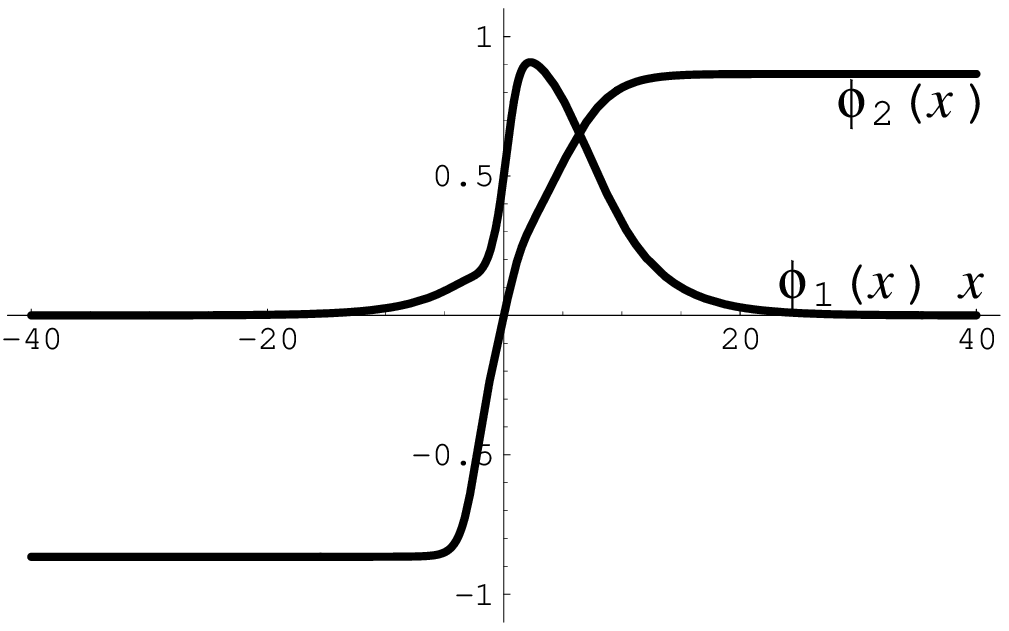,height=1.8cm}
\epsfig{file=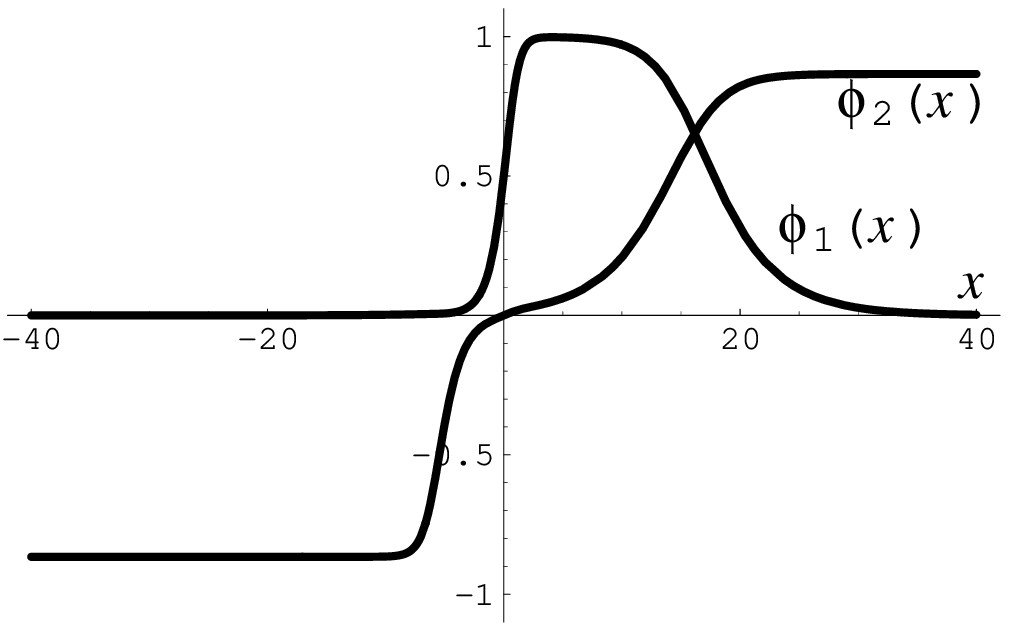,height=1.8cm}} \caption{\it Members of
the family K$_4^{BB}$: from left to right $b=-20$, $-7$, $0$, $7$,
$20$.}
\end{figure}

\begin{figure}[htb]
\centerline{\epsfig{file=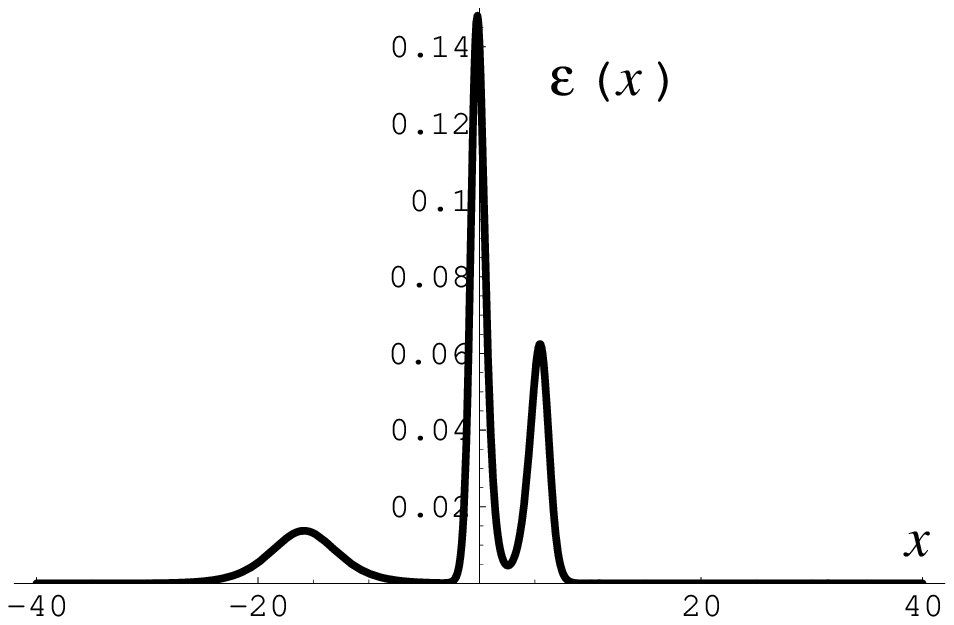,height=1.8cm}
\epsfig{file=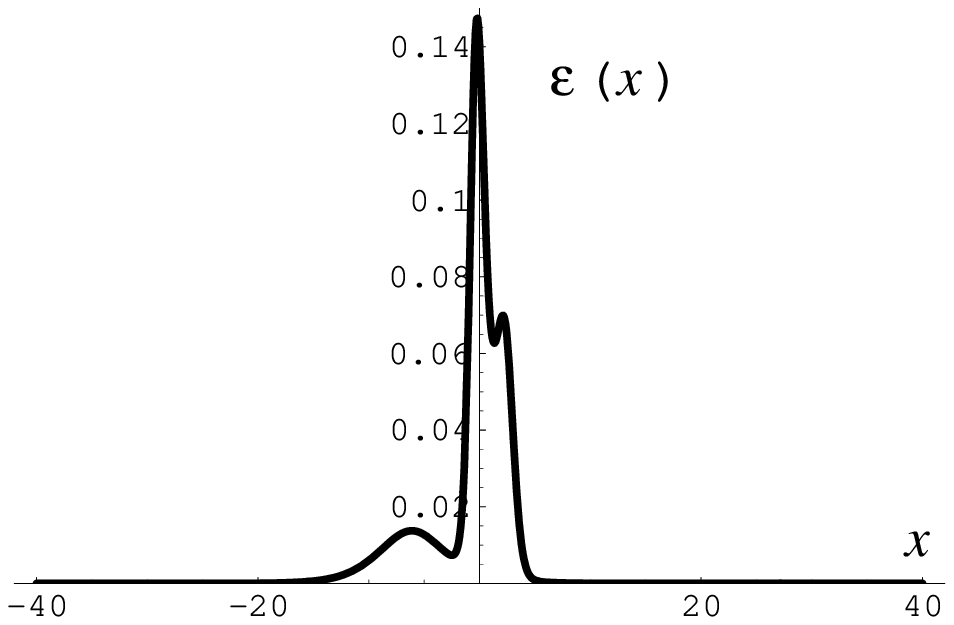,height=1.8cm}
\epsfig{file=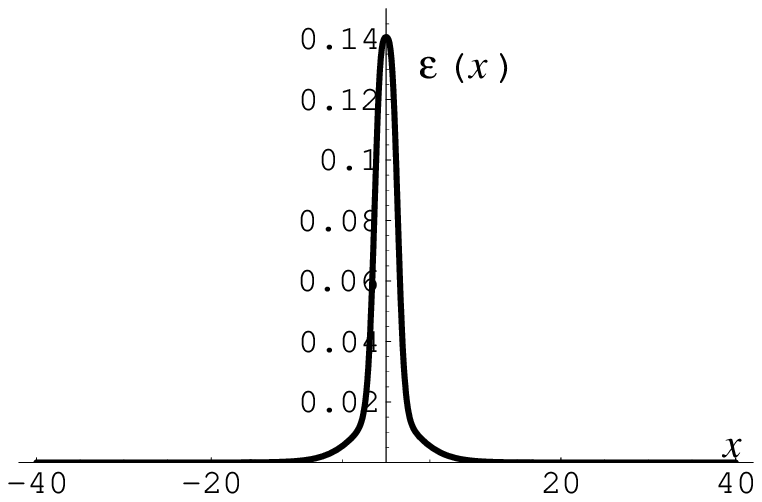,height=1.8cm}
\epsfig{file=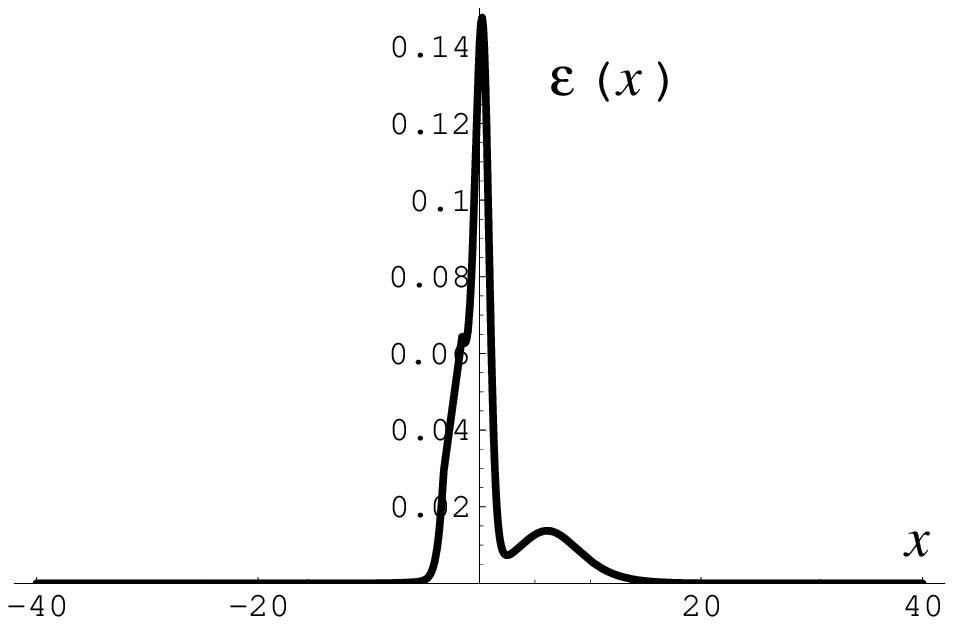,height=1.8cm}
\epsfig{file=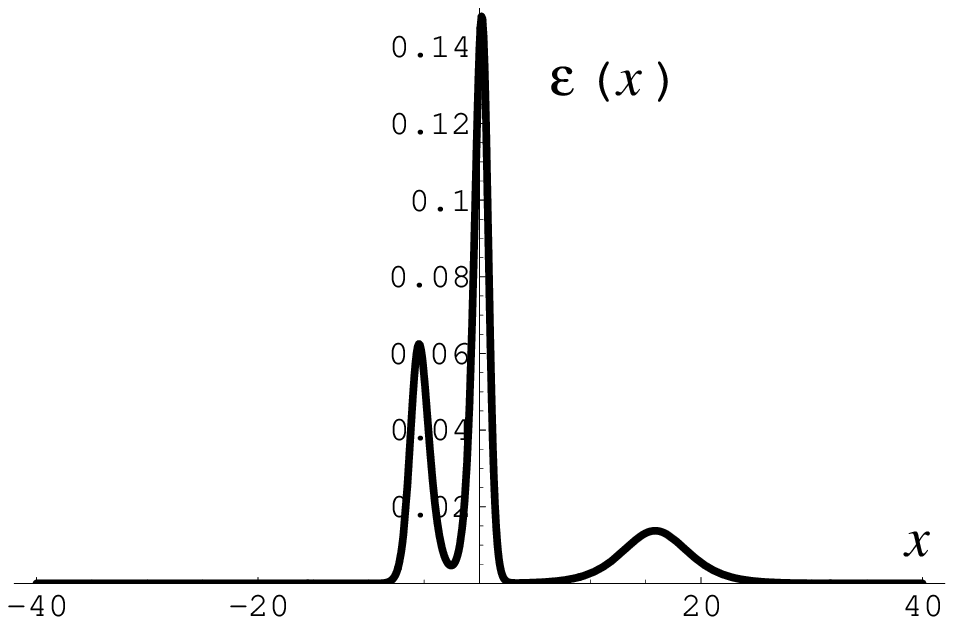,height=1.8cm}} \caption{\it Energy
density of the above kinks: from left to right $b=-20, -7, 0, 7,
20$.}
\end{figure}

In Figures 7 and 8 we respectively plot the kink profiles
(\ref{eq:tk4bb}) and energy density for the $b=-20$, $-7$, $0$,
$7$, $20$ members of this solitary wave family. The main novelty
is that the real component has the shape of a double kink, a fact
clearly perceived from the $b=\pm 20$ plots; the imaginary
component is bell-shaped (slightly distorted). The energy density
graphics show these solitary waves as configurations of three
lumps. We easily identify two of these lumps: the lower lump
corresponds to the basic K$_1^{AB}$ kink whereas the next lump in
height is indeed one of the basic K$_1^{BO}$ kink, compare Figures
3 and 8. The third lump, always in between of the other two, is
precisely the K$_2^{AO}(0)$$\equiv$ K$_2^{AO}$ kink. Taking into
account that the latter are composed of the other two basic kinks,
each member of this family is a configuration of four basic lumps,
two K$_1^{AB}$ and two K$_1^{BO}$ kinks, with the peculiarity that
one K$_1^{AB}$ and one K$_1^{BO}$ are always superposed for any
value of the parameter $b$. In particular, the four lumps coincide
at the same point for the kink configuration characterized by the
value $b=0$. This picture completely explains the intriguing kink
energy sum rule (\ref{eq:srdqen}).

For the $\sigma^2=\frac{1}{2}$ and $b=\pm 1$ values, the solutions
(\ref{eq:tk4bb}) live on semi-circular orbits of radius
$\frac{1}{\sqrt{2}}$ centered at the origin of the internal plane
${\Bbb C}$. The analytical expressions are very simple:
\[
\phi(x)=\frac{1}{\sqrt{2} (1+2 \,e^{(x-a)})} \left[ 2 \sqrt{2}\,
e^{\frac{(x-a)}{2}}+i \left( 1-2 \,e^{(x-a)} \right) \right]\qquad
.
\]

\subsection{Stability and gradient flow lines}

The second-order fluctuation -or Hessian- operator
\[
{\cal H}[\bar{\phi}(x;a,b)]=\left(\begin{array}{cc} -{d^2\over dx^2}+M^2_{11}[\bar{\phi}(x;a,b)] & M^2_{12}[\bar{\phi}(x;a,b)] \\
M^2_{21}[\bar{\phi}(x;a,b)] & -{d^2\over
dx^2}+M^2_{22}[\bar{\phi}(x;a,b)]
\end{array}\right)
\]
valued on a solitary wave solution $\bar\phi(x;a,b)$ is a matrix
differential operator of Schr\"odinger type. Negative eigenvalues
appear in the spectrum of ${\cal H}$ when a solitary wave is
unstable against small fluctuations.

In general the spectral problem of ${\cal H}$ is non-manageable.
There is always, however, important information available: to each
one-parametric family of kink solutions $\bar{\phi}(x,c)$ is
attached the eigenfunction $\frac{\partial\bar{\phi}}{\partial c}$
that belongs to the kernel of ${\cal H}$, i.e.,
$\frac{\partial\bar{\phi}}{\partial c}$ is a zero mode. This is
easily checked by taking the partial derivative of the static
field equations (\ref{eq:seuler}) with respect to $c$.

{\it Stability of K$_1^{BO}$ kinks:} The Hessian operator valued
on the solution (\ref{eq:a5}) is:
\begin{equation}
{\cal H} ({\rm K_1^{BO}})=\left(\begin{array}{cc} {\cal H}_{11} &
{\cal H}_{12} \\ {\cal H}_{21} & {\cal H}_{22}
\end{array}\right)=\left(
\begin{array}{l} -\frac{d^2}{d x^2}-\frac{\sigma^4-6
\sigma^2+1}{4} +(-1)^\beta \frac{\sigma^4-1}{2}\tanh
x+\frac{3\bar\sigma^4}{4} \tanh^2 x
\hspace{1cm} 0 \\[0.2cm] \hspace{1.4cm} 0 \hspace{1.4cm}
-\frac{d^2}{dx^2}+\bar\sigma^4 \left[ -\frac{5}{4} +(-1)^\beta
\frac{3}{2} \tanh x+\frac{15}{4} \tanh^2 x \right]
\end{array} \right) \qquad .\label{eq:hess2}
\end{equation}

In this case, the Hessian reduces to two ordinary Schr\"odinger
operators of Posch-Teller type. ${\cal H}_{11}$ rules the
orthogonal fluctuations to the solution in the internal plane,
whereas ${\cal H}_{22}$ takes into account the tangent
fluctuations to the K$_1^{BO}$ kink. There are no discrete
eigenvalues in the spectrum of ${\cal H}_{11}$. The continuous
spectrum starts at the threshold $\sigma^4$;
$\omega^2(q)=q^2+\sigma^4$ is non-degenerate in the $[\sigma^4,1]$
range but doubly degenerate for eigenvalues higher than 1. ${\cal
H}_{22}$, however, presents a discrete eigenstate with eigenvalue
$\omega^2=0$ (zero mode) and eigenfunction
$\frac{\partial\bar{\phi}}{\partial a}(x,a)$. The continuous
spectrum $\omega^2=q^2+\bar{\sigma}^4$ is non-degenerate for
$\omega^2\in [\bar\sigma^4,4\bar\sigma^4]$ and doubly degenerate
for $\omega^2 \geq 4\bar\sigma^4$. The spectrum of ${\cal
H}(K_1^{BO})$ is semi-definite positive and the K$_1^{BO}$ kinks
are stable. The zero mode identified is associated with
translational invariance; therefore, the perturbation
$\psi_{\omega^2=0}=\frac{\partial\bar{\phi}}{\partial a}(x,a)$ on
the kink causes a infinitesimal translation of the lump.

\vspace{0.2cm}

{\it Stability of the K$_2^{AO}(0)$ kinks:} The Hessian operator
valued on the K$_2^{AO}(0)$ kink is also diagonal:
\begin{equation}
{\cal H} ({\rm K_2}^{AO}(0))=\left(\begin{array}{cc} {\cal H}_{11}
& {\cal H}_{12} \\ {\cal H}_{21} & {\cal H}_{22}
\end{array}\right)=\left( \begin{array}{l}
-\frac{d^2}{dx^2}-\frac{5}{4} +(-1)^\beta \frac{3}{2} \tanh
x+\frac{15}{4} \tanh^2 x
\hspace{2cm} 0 \\[0.2cm] \hspace{0.6cm} 0 \hspace{0.4cm}
-\frac{d^2}{d x^2}-\sigma^2\bar{\sigma}^2-\frac{1}{4} +(-1)^\beta
\left( \sigma^2-\frac{1}{2} \right) \tanh x+\frac{3}{4} \tanh^2 x
\end{array} \right)\qquad
\label{eq:hess3}
\end{equation}
${\cal H}_{11}$, in this case ruling the tangent fluctuations on
the solution, presents a zero mode $\omega^2=0$, with
eigenfunction
\begin{equation}
\psi_{\omega^2=0}^{(1)}= \frac{\partial\bar{\phi}}{\partial
a}(x;a,0)\propto \frac{1}{\sqrt{2 \cosh^3 (x-a) \, e^{(x-a)}}
}\qquad , \label{eq:zerox1}
\end{equation}
associated with translational invariance. The continuous spectrum
shows non-degenerate states in the range $\omega^2\in [1,4]$ and
doubly degenerate states for $\omega^2 \geq 4$. ${\cal H}_{22}$,
determining the behaviour of the orthogonal fluctuations to the
solution, admits another zero mode. The eigenfunction
\begin{equation}
\psi_0^{(2)}=\frac{\partial\bar{\phi}}{\partial b}(x;a,0) \propto
\cosh^{-\frac{1}{2}} (x-a) \, e^{\left( \frac{1}{2} -\sigma^2
\right) (x-a)} \label{eq:zerox2}
\end{equation}
belongs to the kernel of ${\cal H}_{22}$ and exists because the
K$_2^{AO}(0)$ kink is a member of the one-parametric family
K$_2^{AO}(b)$. Applying the perturbation (\ref{eq:zerox2}) to the
solution K$_2^{AO}(0)$ of (\ref{eq:a4}), we obtain
K$_2^{AO}(\delta b)$. The continuous spectrum is non-degenerate in
the range $\omega^2\in [\sigma^4,\bar{\sigma}^4]$ and doubly
degenerate for $\omega^2\geq \bar{\sigma}^4$. Therefore, because
of the positive semi-definiteness of the spectrum of ${\cal H}$ we
conclude that the K$_2^{AO}$ kinks are stable against small
perturbations.

{\it Stability of the K$_2^{AO}(b)$ kinks :} The Hessian for the
other members of the K$_2^{AO}(b)$ family with $b\neq 0$ is not
diagonal. Only the kernel of ${\cal H}$ is known, spanned by the
translational zero mode $\frac{\partial\bar{\phi}}{\partial
a}(x;a,b)$ and the distortional zero mode
$\frac{\partial\bar{\phi}}{\partial b}(x;a,b)$, orthogonal to each
other. Both eigenfunctions obey a neutral equilibrium around a
given kink configuration.

It is possible, however, to conclude that the remaining
eigenvalues are positive by indirect arguments. The solutions to
the first-order system are the flow lines induced by the gradient
of $W^I$. The Hessian matrix
\[
{\rm hess}^I(\bar{\phi})=\left(\begin{array}{cc} \frac{\partial^2
W^I}{\partial\phi_1^2}(\bar{\phi}) & \frac{\partial^2
W^I}{\partial\phi_1\partial\phi_2}(\bar{\phi})\\ \frac{\partial^2
W^I}{\partial\phi_2\partial\phi_1}(\bar{\phi}) & \frac{\partial^2
W^I}{\partial\phi_2^2}(\bar{\phi})
\end{array}\right)
\]
valued at $\bar{\phi}^A$, $\bar{\phi}^B$ and $\bar{\phi}^O$ is
respectively:
\[
{\rm hess}^I(\bar{\phi}^A)=\left(\begin{array}{cc} 1 &
0 \\
 0 & \sigma^2
\end{array}\right)\quad , \quad {\rm hess}^I(\bar{\phi}^B)=\left(\begin{array}{cc} -\sigma^2 &
0 \\
 0 & 2\bar{\sigma}^2
\end{array}\right) \quad , \quad {\rm hess}^I(\bar{\phi}^O)=\left(\begin{array}{cc} -1 &
0 \\
 0 & -\bar{\sigma}^2
\end{array}\right) \quad .
\]
Thus, $\bar{\phi}^A$ is a minimum, $\bar{\phi}^B$ a saddle point
and $\bar{\phi}^O$ a maximum of $W^I$, see Figure 9a. The flow of
${\rm grad}W^I$ runs from $\bar{\phi}^A$ to $\bar{\phi}^O$ along
the non-intersecting orbits (\ref{eq:0k2ao}). There are no
focal/conjugate points and Morse Theory ensures that all these
kinks are stable, see \cite{Ito2} and \cite{J1}.

\begin{figure}[htbp] \centerline{
 \hspace{1cm}\epsfig{file=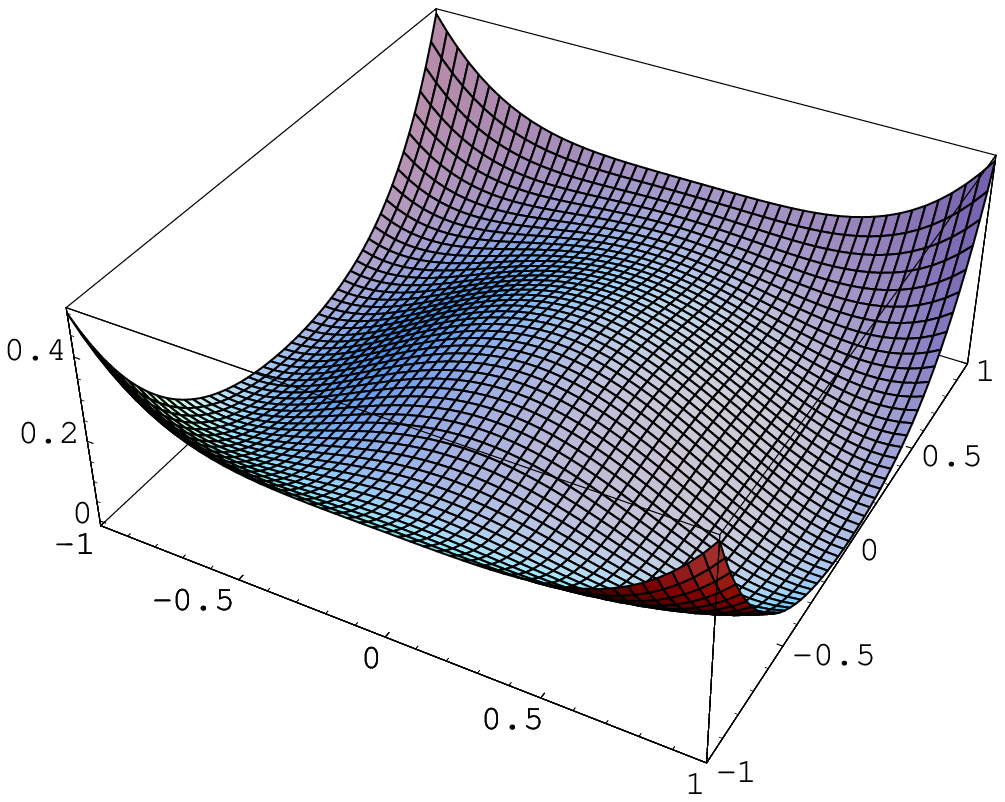,height=5cm} \hspace{1cm}
 \epsfig{file=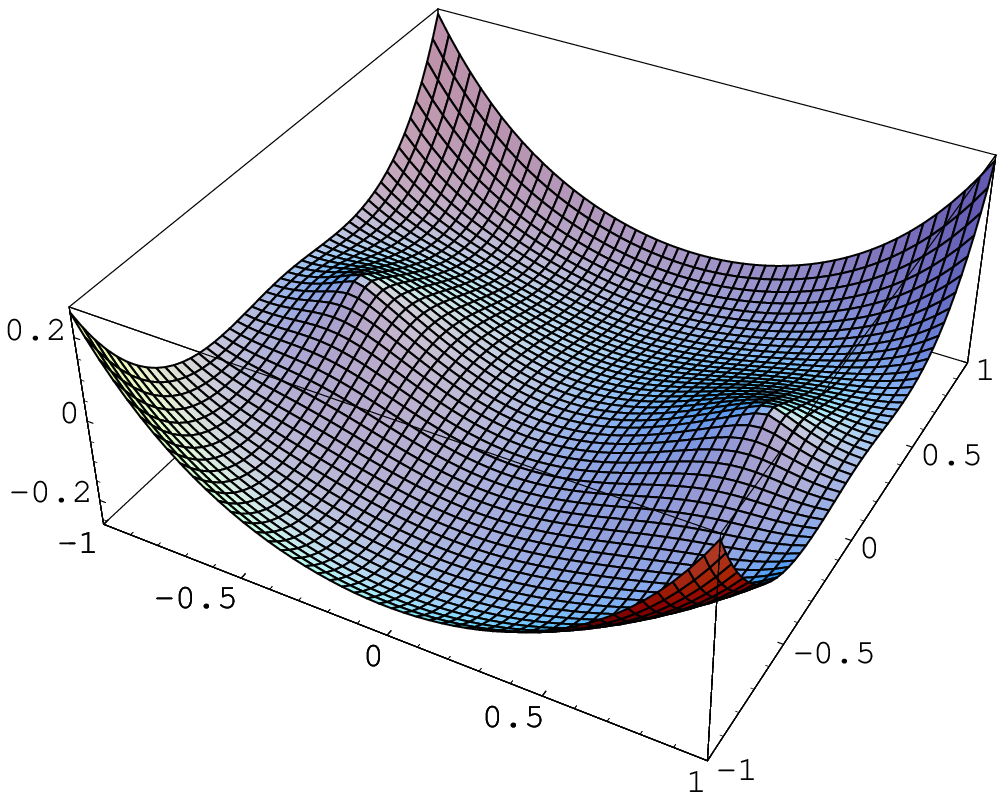,height=5 cm} }
\caption{ \label{super12} a) $W^I(\phi_1,\phi_2)$, $\sigma={1\over
\sqrt{2}}$. \ \ b) $W^{II}(\phi_1,\phi_2) $, $\sigma={1\over
\sqrt{2}}$. }
\end{figure}

{\it Stability of K$_1^{AB}$ kinks:} The K$_1^{AB}$ orbits flow
from a maximum ($\bar{\phi}^A$) to a saddle point ($\bar{\phi}^B$)
of $W^I$. They appear at the $b^2\rightarrow\infty$ limit of the
K$_2^{AO}$ family, no dangerous points are crossed, and the Morse
index theorem tells us that the K$_1^{AB}$ kinks are stable.

{\it Instability of the K$_4^{BB}$(b) kinks:} ${\cal H}$ valued at
the K$_4^{BB}$ kinks is again a non-diagonal matrix Schr\"odinger
operator; there is analytical information available only for the
kernel of ${\cal H}$. The distortional zero mode
$\frac{\partial\bar{\phi}}{\partial b}(x;a,b)$, a Jacobi field in
the language of Variational Calculus, is zero at the foci of the
ellipse for each member of the family (\ref{eq:tk4bb}):
$\frac{\partial\bar{\phi}}{\partial b}(x=a;a,b)=0$. This means
that there exists at least a negative eigenvalue in the spectrum
of ${\cal H}$ because the ground state of (sufficiently
well-behaved) Schr\"odinger operators has no nodes. Therefore, the
K$_4^{BB}(b)$ kinks are unstable. A question arises: what is the
fate of the K$_4^{BB}$ solutions when a perturbation associated
with the negative eigenfunction is exerted? We shall address this
topic later.

A better grasp of the previous result is possible by realizing
that the K$_4^{BB}$ orbits (\ref{eq:horb1}) and (\ref{eq:horb2})
are the flow lines respectively induced by the gradients of
$-W^{II}$ and $W^{II}$. Because
\[
{\rm hess}^{II}(\bar{\phi}^A)=\left(\begin{array}{cc} -2 &
0 \\
 0 & \sigma^2
\end{array}\right)\quad , \quad {\rm hess}^{II}(\bar{\phi}^B)=\left(\begin{array}{cc} \sigma^2 &
0 \\
 0 & 2\bar{\sigma}^2
\end{array}\right) \quad , \quad {\rm hess}^{II}(\bar{\phi}^O)=\left(\begin{array}{cc} 1 &
0 \\
 0 & -\bar{\sigma}^2
\end{array}\right)
\]
$\bar{\phi}^A$ and $\bar{\phi}^O$ are saddle points of $W^{II}$
whereas $\bar{\phi}^B$ is a minimum. The flow lines of $-W^{II}$
accordingly start from $\bar{\phi}^B$ and reach the cuspidal point
$\bar{\phi}^F$ at $x=a$, see Figure 9b. Then, new flow lines, now
induced by $W^{II}$, depart from $\phi^F$ and go to
$\bar{\phi}^B$. $\phi^F$ is a conjugate point to $\bar{\phi}^B$ of
this congruence of trajectories, which are thus unstable according
to Morse Theory.

We end this sub-Section by comment on an intriguing link with
supersymmetry: our bosonic (1+1)-dimensional model admits two
non-equivalent supersymmetric extensions, \cite{AoP}. Acting on
static on shell superfields of the form
\[
\Phi^a[x,\theta]=\phi_a(x)+\bar{\theta}\psi^a(x)-{1\over
2}\bar{\theta}\theta\frac{\partial W}{\partial\phi_a}
\]
one finds two independent sets of ${\cal N}=1$ supercharges:
\begin{eqnarray*}
Q_1^I\phi_a=\theta_2\frac{d\phi_a}{dx}+\psi_1^a-\theta_1\frac{\partial
W^I}{\partial\phi_a} \qquad &,& \qquad
Q_1^{II}\phi_a=\theta_2\frac{d\phi_a}{dx}+\psi_1^a-\theta_1\frac{\partial
W^{II}}{\partial\phi_a} \\
Q_2^I\phi_a=\theta_1\frac{d\phi_a}{dx}+\psi_2^a-\theta_2\frac{\partial
W^I}{\partial\phi_a} \qquad &,& \qquad
Q_2^{II}\phi_a=\theta_1\frac{d\phi_a}{dx}+\psi_2^a-\theta_2\frac{\partial
W^{II}}{\partial\phi_a} \qquad .
\end{eqnarray*}
Here, $\theta=\left(\begin{array}{c}\theta_1 \\
\theta_2\end{array}\right)$ are Grassman Majorana spinors that
span the odd part of superspace; $\psi^a=\left(\begin{array}{c}\psi^a_1 \\
\psi^a_2\end{array}\right)$ are the two Majorana spinor fields,
superpartners of the bosonic fields. Double solitary waves are
classical BPS states, henceforth stable, because they are
annihilated by a combination of the $Q^I$ supercharges:
\[
Q^I_+\phi_a=(Q^I_1+Q_2^I)\phi_a=(\theta_1+\theta_2)\frac{d\phi_a}{dx}
+(\psi_1^a+\psi_2^a)-(\theta_1+\theta_2)\frac{\partial
W^{I}}{\partial\phi_a}=0
\]
if, moreover, $\psi_1^a+\psi_2^a=0$. Quadruple solitary waves,
however, are non BPS because $Q_+^{II}\phi^a\neq 0$ for them. As a
curious fact, despite of having two sets of ${\cal N}=1$
supercharges this model does not admit ${\cal N}=2$ supersymmetry
because $W(\phi,\bar{\phi})=W^I(\phi_a)+iW^{II}(\phi_a)$ is non
holomorphic.

\section{Adiabatic motion of non-linear waves}

In sum, the K$_2^{AO}(a,b)$ and K$_4^{BB}(a,b)$ solitary wave
families depend on $a\in{\Bbb R}$, a translational parameter
setting the center of the kink, and $b\in{\Bbb R}$, the parameter
measuring the distortion of the kink shape from a single lump. The
space of kink solutions is thus the $(a,b)\in{\Bbb R}^2$-plane,
with the basic kinks living in the boundary circle $\partial{\Bbb
R}^2\simeq{\Bbb S}^1_\infty$ at infinity. In fact, the moduli
space of double solitary waves is only the upper half-plane
because invariance under $\bar{\phi}_2\rightarrow -\bar{\phi}_2$
comes from invariance under $b\rightarrow -b$ for the K$_2^{AO}$
kinks.

We now ask the following question: Are there solitary wave
solutions to the full wave equations
(\ref{eq:euler1})-(\ref{eq:euler2}) such that changes in shape
take place? For normal kinks, the dynamics is merely dictated by
Lorentz invariance and thus characterized by shape invariance. We
shall address the difficult issue of the evolution of composite
solitary waves within the framework of Manton's adiabatic
principle, see \cite{Manton} and \cite{Mar}: geodesics in the
moduli space determine the slow motion of topological defects. In
this scheme, the dynamics arises from the hypothesis that only the
parameters of the moduli space $a$ and $b$ depend on time.
Moreover, the kink evolution $\bar{\phi}^K(x;a(t),b(t))$ is slow
enough to guarantee that the static differential equations
(\ref{eq:seuler}) will be satisfied for every given $t$ in a good
approximation. Therefore, the field theoretical action
(\ref{eq:act}) reduces to
\[
S^G=\frac{1}{2}\int dt \sum_{i=1}^2 \frac{\partial
\bar{\phi}^K_i}{\partial t} \frac{\partial\bar{\phi}^K_i}{\partial
t} =\int dt \left[
\frac{1}{2}g_{aa}(a,b)\frac{da}{dt}\frac{da}{dt}+g_{ab}(a,b)\frac{da}{dt}\frac{db}{dt}+
\frac{1}{2}g_{bb}(a,b)\frac{db}{dt}\frac{db}{dt} \right]\quad ,
\]
where
\[
g_{aa}(a,b)= \sum_{i=1}^2 \int_{-\infty}^\infty\, dx
\frac{\partial\bar{\phi}^K_i}{\partial a}\frac{\partial
\bar{\phi}^K_i}{\partial a}\quad , \quad  g_{bb}(x,b)=
\sum_{i=1}^2 \int_{-\infty}^\infty \, dx \frac{\partial
\bar{\phi}^K_i}{\partial b}\frac{\partial\bar{\phi}^K_i}{\partial
b}\quad ,
\]
\[
g_{ab}(a,b)= \sum_{i=1}^2 \int_{-\infty}^\infty\, dx
\frac{\partial \bar{\phi}^K_i}{\partial a}\frac{\partial
\bar{\phi}^K_i}{\partial b} \quad ,
\]
will be understood as the components of a metric tensor.

We think of $S^G$ as the action for geodesic motion in the kink
space with a metric inherited from the dynamics of the zero modes.
The geodesic equations coming from the action $S^G$ determine the
behaviour of the variables $a(t)$ and $b(t)$ and therefore
describe the evolution of the lumps $\bar{\phi}^K(x;a(t),b(t))$
within the configurations of a specified kink family.

\subsection{Adiabatic evolution of double solitary waves}

The double solitary waves K$_2^{AO}(a,b)$ can be seen as two
particles moving on a line. Geodesic motion in the $(a,b)$-plane
would correspond to the slow-speed dynamics of the the
two-particle system described in terms of the center of mass and
relative coordinates $a$ and $b$.

For the value $\sigma=\frac{1}{\sqrt{2}}$, the formula
(\ref{eq:tk2ao}) describing the K$_2^{AO}$ kinks is simpler and
the tensor metric can be written explicitly. By changing variables
to $z=e^x$, the integrals become of rational type, and we obtain:
\begin{eqnarray}
g_{aa}(a,b)=\frac{1}{4}\quad &,& \quad g_{ab}(a,b)={b\over 4}
g_{bb}(a,b)=g_{ba}(a,b) \nonumber\\
g_{bb}(a,b)&=&\frac{1}{(1-b^4)^\frac{3}{2}}\left( {\rm
arccot}\,\frac{b^2}{\sqrt{1-b^4}}-b^2 \sqrt{1-b^4} \right) \quad .
\label{eq:met2}
\end{eqnarray}
From the graphics in Figure 10 we see that the metric induced in
the $(a,b)$-plane by the zero mode dynamics has good properties:
$g_{aa}>0$, $g_{bb}>0$, and, $g=g_{aa}g_{bb}-g_{ab}^2>0$.

\begin{figure}[htbp]
\centerline{\epsfig{file=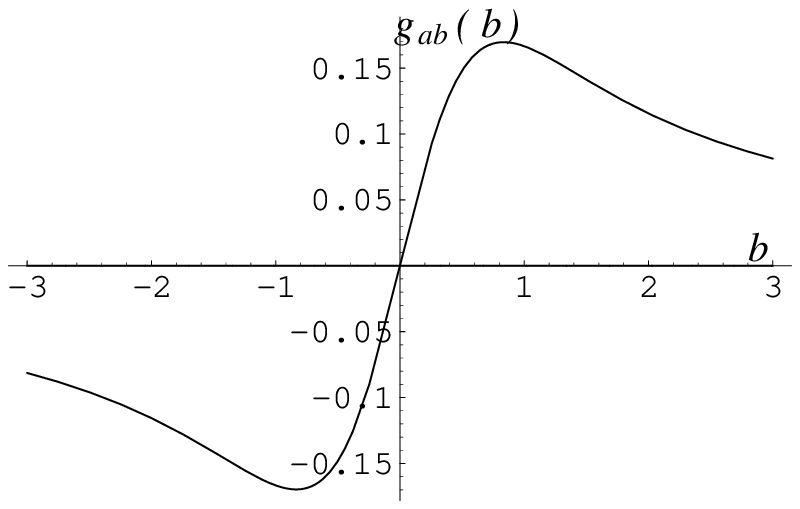,height=2.5cm}\hspace{0.2cm}
\epsfig{file=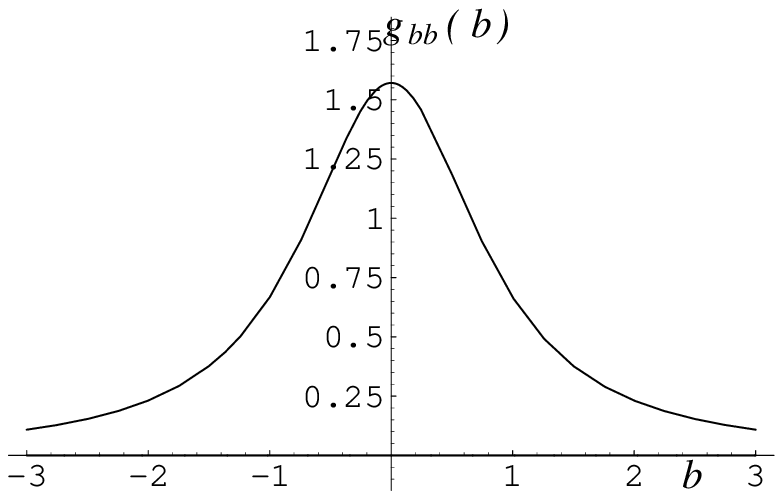,height=2.5cm}
\hspace{0.2cm}\epsfig{file=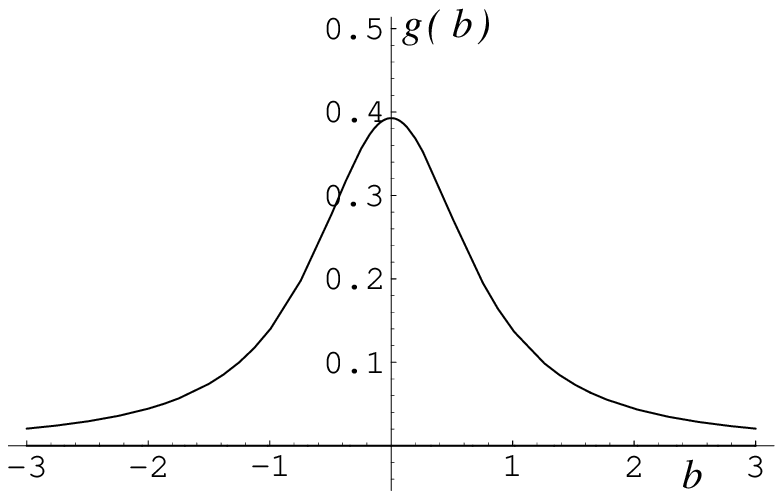,height=2.5cm}}
\caption[Figura]{\small Graphics from left to right of $g_{ab}$,
$g_{bb}$ and $g=g_{aa} g_{bb}-g_{ab}^2$ as functions of $b$.}
\end{figure}

In fact, metrics of the general form
\[
ds^2={1\over 4}da^2+2F(b)dadb+G(b)db^2 \qquad ,
\]
with both $G(b)$ and $g(b)={1\over 4}G(b)-F^2(b)$ greater than
zero for all $b$, are always flat. Only two Christoffel symbols
are non-null:
\[
\Gamma^a_{bb}={1\over 2g(b)}\left(2G(b){dF\over db}-F(b){dG\over
db}\right) \qquad , \qquad \Gamma^b_{bb}= -{1\over
8g(b)}\left(8F(b){dF\over db}-{dG\over db} \right) \qquad .
\]
Accordingly, the only relevant component of the curvature tensor
is zero: $R^a_{\,bab}=0$. It is therefore possible to write the
metric in the form
\[
ds^2=d\tilde{a}^2+d\tilde{b}^2
\]
by means of the isometric transformation
$(\tilde{a}(a,b),\tilde{b}(a,b))$ if the new coordinates satisfy
the PDE system:
\begin{equation}
\left(\frac{\partial\tilde{a}}{\partial
a}\right)^2+\left(\frac{\partial\tilde{b}}{\partial
a}\right)^2={1\over 4} \quad , \quad
\frac{\partial\tilde{a}}{\partial
a}\cdot\frac{\partial\tilde{a}}{\partial
b}+\frac{\partial\tilde{b}}{\partial
a}\cdot\frac{\partial\tilde{b}}{\partial b}=F(b) \quad , \quad
\left(\frac{\partial\tilde{a}}{\partial
b}\right)^2+\left(\frac{\partial\tilde{b}}{\partial
b}\right)^2=G(b) \qquad . \label{eq:isom}
\end{equation}
Zero curvature ensures the integrability of (\ref{eq:isom}) and
solutions for $(\tilde{a},\tilde{b})$ can be found. In particular,
restriction to the form $(\tilde{a}={a\over
2\sqrt{2}}+\bar{a}(b),\tilde{b}={a\over 2\sqrt{2}}+\bar{b}(b))$
provides the solution:
\[
\bar{a}=\sqrt{2}\int_0^b \, db^\prime \, [F(b^\prime)\pm
\sqrt{G(b^\prime)}] \qquad , \qquad \bar{b}=\sqrt{2}\int_0^b \,
db^\prime \, [F(b^\prime)\mp \sqrt{G(b^\prime)}] \qquad .
\]
Geodesics are thus straight lines in the $(\tilde{a},\tilde{b})$
variables:
\[
\tilde{a}(t)=a_1 t+a_2 \qquad \qquad , \qquad \qquad
\tilde{b}=b_1 t+b_2 \qquad ,
\]
where $a_1 , a_2 , b_1 , b_2$ are integration constants.

\begin{figure}[htb]
\centerline{\epsfig{file=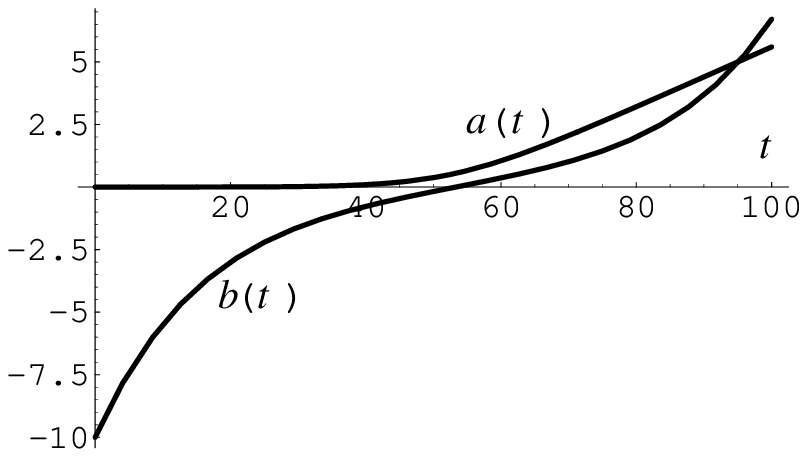,height=3.5cm}
\hspace{1cm} \epsfig{file=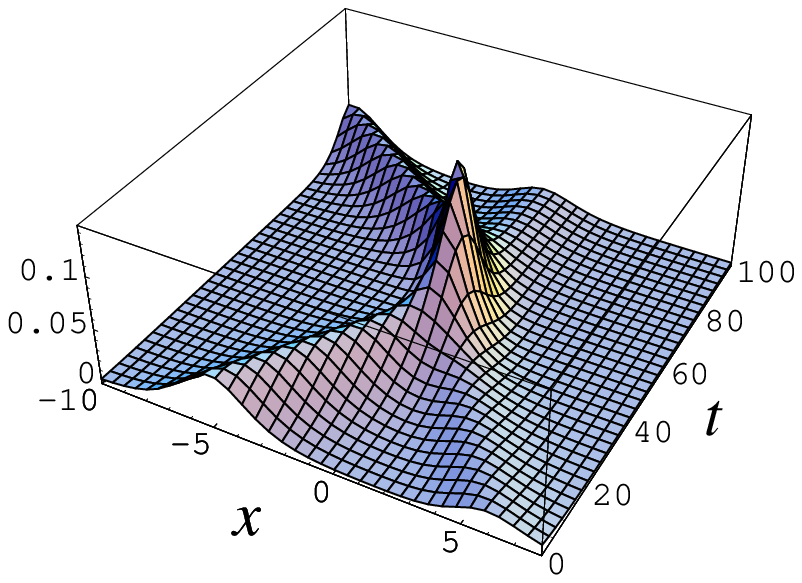,height=4.5cm}}
\caption{\it Evolution of the moduli variables and evolution of
the lumps in the sector AO.}
\end{figure}

Back in the original coordinates, a typical $(a(t),b(t))$ geodesic
is depicted in Figure 11 (left). The $b(t)$ curve shows that the
relative distance between the two basic kinks decreases with time
until they become fused -$b=0$- at $t=50$. Then, the distance
starts to grow towards a complete splitting of the K$_1^{AB}$ and
K$_1^{BO}$ kinks in the remote future. The evolution of the center
of mass is unveiled by the $a(t)$ curve: it remains at $x=0$ while
the two basic kinks are far apart. Shortly before the collision,
$a$ starts to grow and thereafter continuous to increase. The two
kinks rebound, the center of mass being closer to the heaviest
kink. The picture is clearer in the plot of the energy density
evolution along this geodesic shown in Figure 11 (right). Shortly
after $t=-\infty$, the geodesic starts from a point in the moduli
space corresponding to one K$_1^{AB}$ kink and one K$_1^{BO}$ kink
very far apart. The two basic lumps begin to approach each other
as $|b|$ decreases, distorting their shapes when they  start to
collide. Later, a single K$_2^{AO}(0)$ lump is formed when $b(t)$
reaches $0$ (see Figures 5 and 11 (right)). At this point, the two
lumps bounce back following the above-described motion in reverse,
i.e., the separation between the two lumps increases, running
asymptotically towards the configuration of two infinitely
separated kinks in the ${\cal C}^{AO}_{\pm}$ sector. Thinking of
the graphics in Figure 5 as the stills of a movie, the adiabatic
dynamics of the double solitary waves prescribes the speed at
which the film is played.

\subsection{Meta-stable slow motion of quadruple solitary waves}
We now perform the same analysis in the ${\cal C}^{BB}_{\pm\mp}$
sectors, again for the value $\sigma=\frac{1}{\sqrt{2}}$. The
metric
\begin{eqnarray*}
&&
h_{aa}(a,b)=2h_{bb}(a,b)=\frac{1}{2}\\
h_{ab}(a,b)&=&-\frac{1}{32} \left[ 10+
\frac{e^b}{(e^b-1)^\frac{3}{2}} \left(\pi {\rm Sign}(e^b-1) - 2
{\rm arctan}\frac{1}{\sqrt{e^b-1}}
\right) \right. \\
 && \left. - \frac{\sqrt{e^b}}{(1-e^b)^\frac{3}{2}}
\left(- \pi {\rm Sign}(1-e^b) + 2{\rm
arctan}\frac{\sqrt{e^b}}{\sqrt{1-e^b}}
 \right) \right]
\end{eqnarray*}
is of the general form studied in the previous sub-Section.
\begin{figure}[htbp] \centerline{
 \hspace{1cm}\epsfig{file=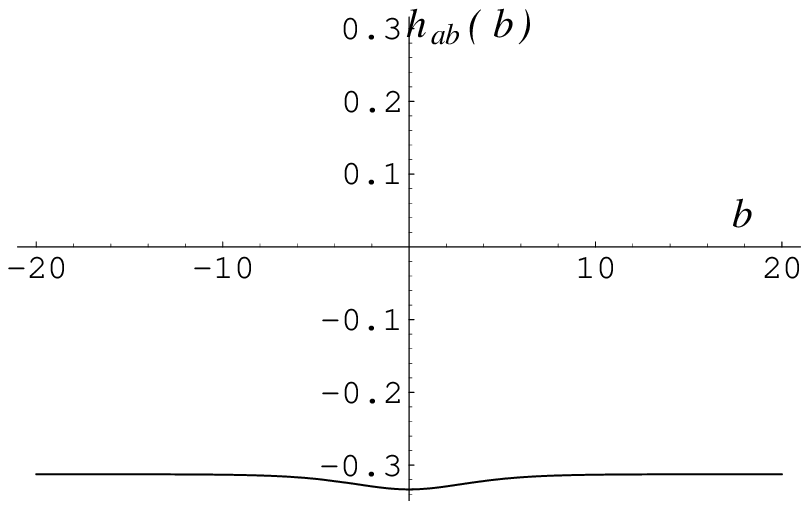,height=2.5cm} \hspace{1cm}
 \epsfig{file=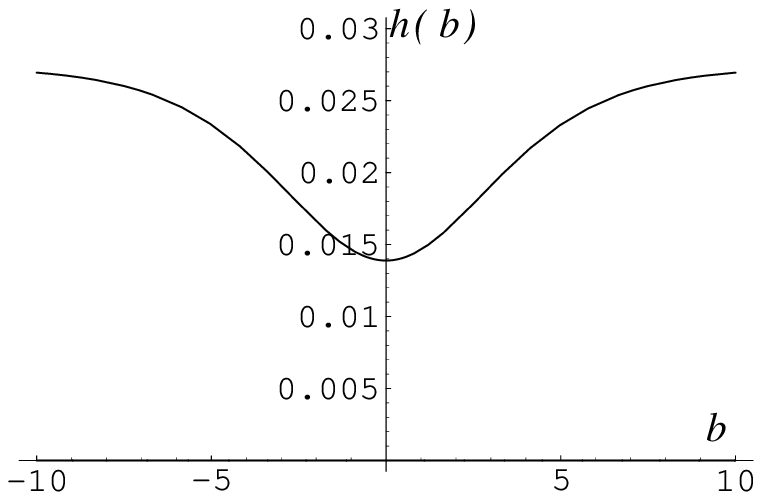,
 height=2.5cm} } \caption[Figura] {\small Graphics from left to right of $h_{ab}$
and $h=h_{aa} h_{bb}-h_{ab}^2$ as functions of $b$.}
\end{figure}

The geodesics are easily found and plotted using Mathematica in
Figure 13 (left). The evolution of the energy density along this
geodesic is also shown in Figure 13 (right). The motion begins at
$t=-\infty$ with three lumps very far away from each other along
the real line , arranged as a K$_1^{BO}$, K$_2^{AO}$, K$_1^{AB}$
configuration, ordered by increasing $x$. Then, the lumps start to
approach the K$_2^{AO}$ kink until they coincide, forming the
K$_4^{BB}(0)$ configuration. Later, the three lumps start to split
in such a way that the K$_1^{BO}$ kink overtakes the K$_2^{AO}$
and K$_1^{AB}$ lumps whereas the latter kink is also passed by the
K$_2^{AO}$ kink. This motion naturally proceeds towards
K$_1^{AB}$, K$_2^{AO}$, K$_1^{BO}$ configurations with growing
distances between the basic lumps. Note that there are no problems
in this sector at $b=\pm\infty$; contrarily to the $g$-metric the
$h$-metric does not vanish at infinity. This means that some lumps
could disappear smoothly as an effect of this low-energy
scattering.

\begin{figure}[htb]
\centerline{ \epsfig{file=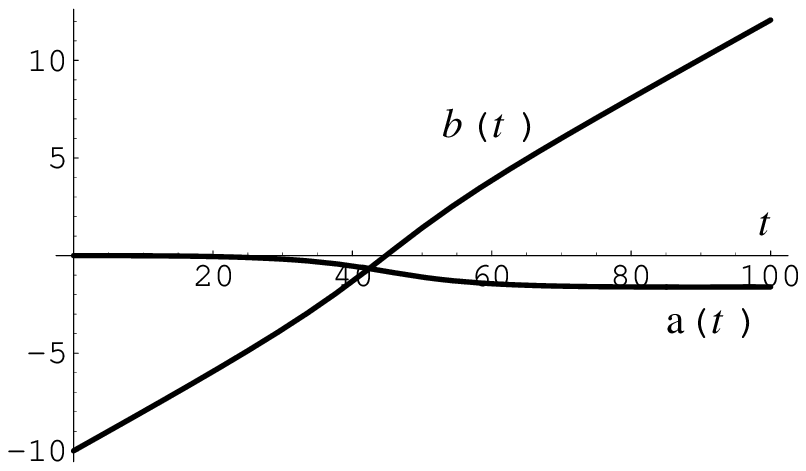,height=3.5cm}
\hspace{1cm} \epsfig{file=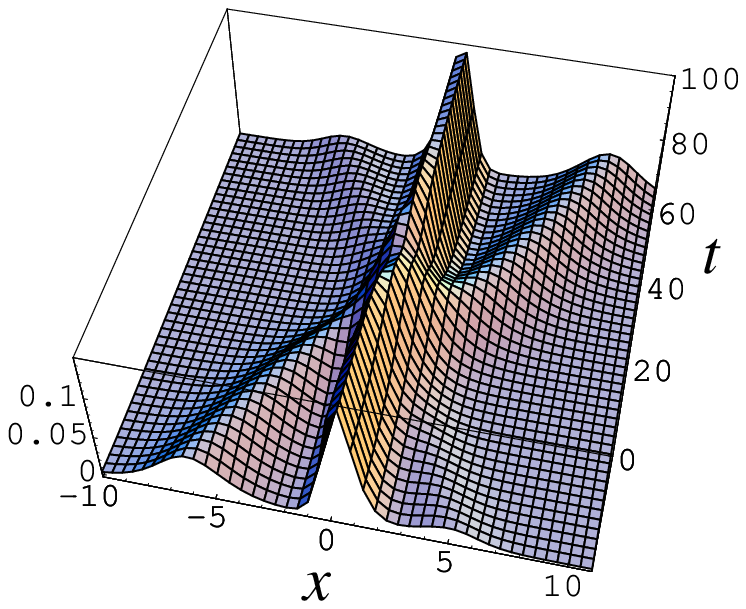,height=4.5cm}}
\caption{\it Evolution of the moduli variables and evolution of
the lumps in the sector ${\cal C}^{BB}_{\pm\mp}$.}
\end{figure}

In fact, in Section \S  3 we argued that the K$_4^{BB}$ solitary
waves are unstable by demonstrating the existence of a negative
eigenfunction of the Hessian. Fluctuations in this direction of
the configuration space will destroy the gentle evolution between
the solitary waves described above. Closer understanding of the
non-adiabatic process triggered by the negative fluctuation is
possible by starting from a K$_4^{BB}(b)$ lump with large $b$:
there are one K$_1^{AB}$ and one K$_1^{BO}$ lumps, respectively
near $x=-\infty$ and $x=+\infty$, together with a K$_2^{AO}$ kink
sitting at the origin $x=0$. The zero mode
\[
\psi_{\omega^2=0}^{(2)}=\frac{1}{\sqrt{{\rm cosh}(x-a)}}
\]
of the Hessian at K$_2^{AO}(0)$ provides a very good approximation
to the negative eigenfunction of the Hessian at
K$_4^{BB}(b\rightarrow\infty)$ because this latter solitary wave
tends asymptotically to K$_2^{AO}(0)$, $\psi^{(0)}$ has no nodes,
and the topological barriers in the ${\cal C}^{AO}_\pm$ and ${\cal
C}^{BB}_{\pm\mp}$ sectors are different. Moreover, the
$\psi^{(2)}_0$ fluctuation causes the splitting of the K$_2^{AO}$
lump and the evolution reveals the motion of four lumps: two
K$_1^{AB}$ and two K$_1^{BO}$ kinks. Of course, this configuration
does not belong to the static kink family (\ref{eq:tk4bb}); the
negative fluctuation suppresses the overlapping of two basic
lumps.

This picture is confirmed by solving via numerical analysis the
Cauchy problem for the PDE system
(\ref{eq:euler1})-(\ref{eq:euler2}) with initial conditions
\[
\phi(x,0)={\rm K}_4^{AB}(30)(x) \qquad \qquad , \qquad \qquad
\frac{\partial\phi}{\partial t}(x,0)=\varepsilon\psi^{(2)}_0(x)
\]
and Dirichlet boundary conditions, see Figure 14. It seems that
the two K$_1^{BO}$ attract each other but there is repulsion
between the two K$_1^{AB}$ kinks.

\begin{figure}[htb]
\centerline{ \epsfig{file=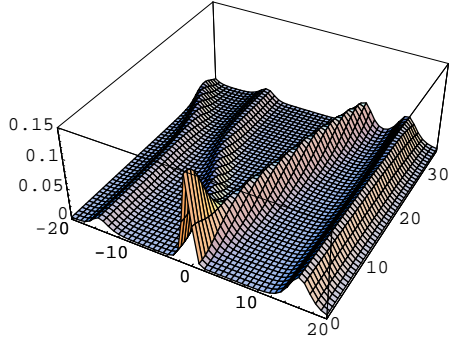,height=4.5cm} }
\caption{\it Evolution of a ${\rm K_4}^{BB}(b)$ solitary wave
after $\psi^{(2)}_0$ fluctuations.}
\end{figure}

\subsection{Perturbing the system: non-linear wave bound state}

Finally, we add an small perturbation to our system. The basic
kinks K$_1^{BO}$ and K$_1^{AB}$ move adiabatically along the
geodesics of the metric induced by the kinetic energy on the kink
moduli space in the ${\cal C}^{AO}_\pm$ sector. We choose to
induce forces between the basic kinks by the following
modification of $V(\phi_1,\phi_2)$:
\[
V'(\phi_1,\phi_2)=V(\phi_1,\phi_2)+\varepsilon v(\phi_1,\phi_2)
\qquad , \qquad  v(\phi_1,\phi_2)=\phi_1^2\left(
\phi_1^2+\frac{\phi_2^2}{1-\sigma^2}-1\right)^2 \qquad .
\]
Here, $\varepsilon$ is an small parameter and our choice is such
that the zeroes of $V$ are also zeroes of $V'$; thus, $V'$ also
comply with Coleman's theorem and enters into the class of
admissible deformations of $V_{\rm CSH}$ in (1+1)-dimensions.
Moreover, the K$_1^{AB}$ and K$_1^{BO}$ kinks survive this
perturbation without any changes: they are still solitary wave
solutions with the same energy. The K$_2^{AO}$ solitary wave,
however, needs to be adjusted:
\[
\bar{\phi}(x,a)=(-1)^\alpha{1\over\sqrt{2}}\sqrt{1+(-1)^\beta{\rm
tanh}[\sqrt{1+2\varepsilon}(x-a)]} \qquad , \quad
 \alpha,\beta=0,1 \qquad ,
\]
is a solitary wave solution of the perturbed system with higher
energy: ${\cal E}'({\rm K_2}^{AO})={1\over
4}\sqrt{1+2\varepsilon}$. The perturbation is chosen in such a way
that the kink energy sum rule (\ref{eq:sumrk2}) is broken.

All the other solitary wave are destroyed by the perturbation.
Nevertheless, the adiabatic approximation can still be used to
describe the slow motion of the basic lumps. For small enough
$\varepsilon$, $\varepsilon << 1$, the solitary waves
(\ref{eq:tk2ao}) are approximate solutions of the perturbed
system; therefore, we take the parameters $a$ and $b$ as the
moduli variables, even though the perturbation forbids neutral
equilibria in the moduli space. Plugging the approximate solutions
into the action of the perturbed model we obtain:
\begin{equation}
S^G=\int dt \left[\frac{1}{2}g_{aa}\frac{da}{dt}
\frac{da}{dt}+g_{ab}\frac{da}{dt}
\frac{db}{dt}+\frac{1}{2}g_{bb}\frac{db}{dt}
\frac{db}{dt}-\varepsilon u(b) \right]\qquad . \label{eq:sfin}
\end{equation}
The perturbation induces a potential energy, plotted in Figure 15
(left),
\[
u(b)=\int_{-\infty}^\infty \, dx \, v({\rm K_2}^{AO}(b)(x))
\]
and the motion ceases to be geodesic; $u(b)$ is a potential
barrier with maximum height reached at $b=0$ that goes to zero for
$b=\pm\infty$. The equations of motion coming from (\ref{eq:sfin})
determine the evolution of the moduli variables $a(t)$ and $b(t)$
with time.

In Figure 15 (center) the plot of both $a$ and $b$ as a function
of time is shown for a solution of the equations of motion with
$\varepsilon=0.1$ and initial conditions $a(0)=0$, $b(0)= -2.004$,
$a^\prime (0)=0$, $b^\prime (0)=0.001$. The surprising outcome is
that $b$ seems to be quasi-periodic in time; this is a Sisyphean
movement, crossing over and over again the top of the hill!
{\footnote{This bizarre behavior was first discovered in Kapitsa
oscillator, see \cite{Kap}.}}

In Figure 15 (right) the evolution of the energy density along
this quasi-periodic orbit is drawn. We observe that the net effect
of the perturbation plus the non-trivial metric is the upsurge of
an attractive force between the two basic lumps K$_1^{BO}$ and
K$_1^{AB}$, giving rise to a bound state similar to the breather
mode in the sine-Gordon model.

\begin{figure}[htb]
\centerline{\epsfig{file=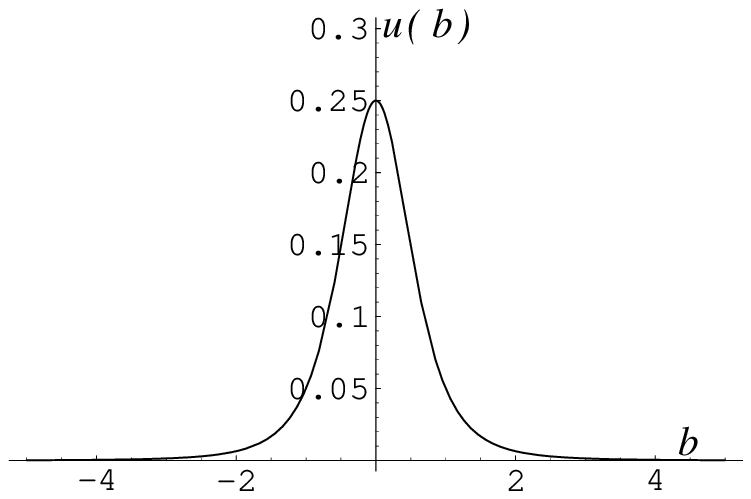,height=2.5cm}
\hspace{0.1cm} \epsfig{file=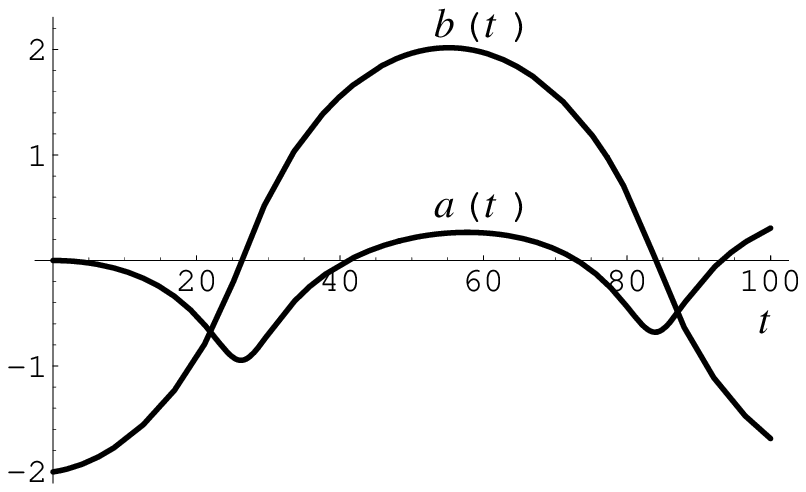,height=2.5cm}
\hspace{0.3cm} \epsfig{file=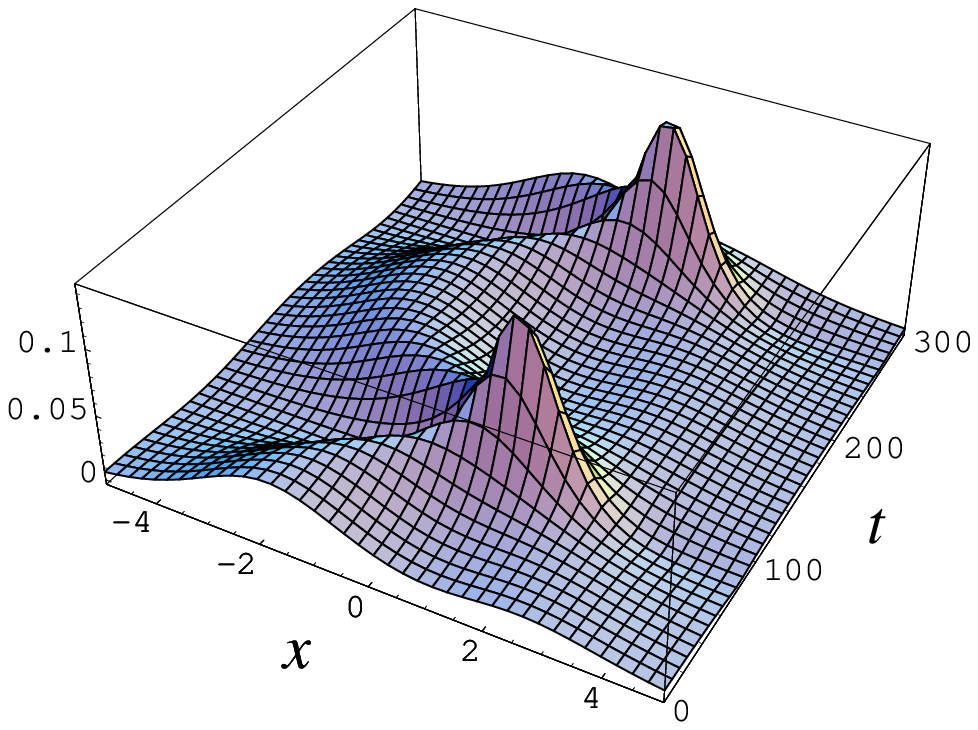,height=3.5cm}}
\caption{\it Plot of $u(b)$, evolution of the moduli variables and
evolution of the lumps in the deformed model.}
\end{figure}

\section*{Acknowledgements} One of us, JMG, thanks A. Chirsov and
his collaborators from Sankt Petersburg University for showing him
an experimental demonstration of the Kapitsa oscillator.

\end{document}